\documentclass[10pt,twocolumn]{article}
\usepackage{latex8}

\usepackage{latexsym,color,graphics}
\usepackage{diagrams}
\usepackage{url}
\usepackage{epsfig}
\usepackage{amssymb}
\usepackage{amsmath}
\usepackage{amsfonts}

\begingroup
\catcode`\~=11
\gdef\urltilde{\lower 0.6ex\hbox{~}}
\endgroup

\newcommand{\A}{\mathcal{A}} \newcommand{\B}{\mathcal{B}}

\newcommand{\G}{\mathcal{G}} 
 
\newcommand{\K}{\mathcal{K}} \renewcommand{\L}{\mathcal{L}}
\newcommand{\M}{\mathcal{M}} 
\renewcommand{\O}{\mathcal{O}} \renewcommand{\P}{\mathcal{P}}
 \newcommand{\R}{\mathcal{R}}

\newtheorem{theo}{Theorem}

\newtheorem{propo}{Proposition}

\newtheorem{coro}{Corollary}
\newtheorem{definition}{Definition}

\begin{document}

\title{ A New Representation Theorem for Many-valued Modal Logics}


\author{Zoran Majki\'c\\
 International Society for Research in Science and Technology,\\
 PO Box 2464 Tallahassee, FL 32316 - 2464 USA\\
 http://zoranmajkic.webs.com/
 \email{majk.1234@yahoo.com}
}


\maketitle \thispagestyle{empty}
%
\begin{abstract}
We propose a new definition of the representation theorem for
many-valued   logics,  with modal operators as well, and define the
stronger relationship between algebraic models of a given logic  and
relational structures used to define the Kripke possible-world
semantics for it. Such a new framework offers  a new semantics for
many-valued logics
based on the truth-invariance entailment. Consequently, it is
substantially different from current definitions based on
 a matrix with a designated subset
of logic values,  used for the satisfaction relation, often
difficult to fix. In the case when the many-valued modal logics are
based on the set of truth-values that are complete distributive
lattices we obtain a compact autoreferential Kripke-style canonical
representation.  The Kripke-style semantics for this subclass of
modal logics have the joint-irreducible subset of the carrier set of
many-valued algebras as set of possible worlds. A significant member
of this subclass is the paraconsistent fuzzy logic extended by new
logic values in order to also deal with incomplete and inconsistent
information. This new theory is applied for the case of
autoepistemic intuitionistic many-valued logic, based on Belnap's
4-valued bilattice, as a minimal extension of classical logic used
to manage incomplete and inconsistent information as well.
\end{abstract}


%

%

\section{Introduction}
Many-valued logic was conceived as a logic for uncertain, incomplete
and possibly inconsistent information which is very close to the
statements containing the words "necessary" and "possible", that is,
to the statements that make an assertion about the \emph{mode of
truth} of some other statement. \emph{Algebraic} semantics
interprets modal connectives as operators,
 while \emph{Relational} semantics uses
relational structures, often called Kripke models, whose elements
are thought of variously as being possible worlds; for example,
moments of time, belief situations, states of a computer, etc.. The
two approaches are closely related: the subsets of relational
structures form an algebra with modal operators, while conversely
any modal algebra can be embedded into an algebra of subsets of a
relational structure via extensions of Stone's Boolean
representation theory. For example, the first (1934) and the most
known \emph{Stone's representation theorem} for Boolean algebras
\cite{John82},  is the duality between the category of Boolean
algebras and the category of Stone spaces. Every Boolean algebra
$(BA,+,\cdot, \setminus, 0,1)$, where $+,\cdot,\setminus$ are
corresponding algebraic operations (addition, multiplication and
complement) for classical logic connectives $\vee, \wedge, \neg$
respectively, is isomorphic to an algebra of particular clopen
(i.e., simultaneously closed and open) subsets of its Stone space.
Stone's theorem has since been the model for many other similar
representation theorems. Our representation theorem, in the case of
distributive complete lattice of truth values, is a particular
Stone-like autoreferential representation
based on the particular subsets of these truth-values.\\
In order to be able to follow this paper  the readers must have
clear in mind the difference between a many-valued \emph{logic} and
its underlying \emph{algebra} of truth-values (for example, the
propositional logic and its Boolean algebra, the intuitionistic
logic and its Heyting algebra), so that we can informally use the
term lattice (of algebraic truth values) speaking about logics as
well.\\ Given two sets $A$ and $B$, we  denote by  $A^B$, the set of
all functions from $B$ to $A$,  by $A^n$ the n-th cartesian product
$A \times ...\times A$, and by $\P(A)$ the powerset of $A$.\\
The representation theorems are based on Lindenbaum algebra of a
logic $\L = (Var, \O, \Vvdash)$, where $Var$ is a set of
propositional symbols of a language $\L$, $\O$ is the set of logical
connectives and $\Vvdash$ is the entailment relation of this logic.
We denote by $F(\L)$ the set of all formulae. Notice that the
 truth-values in $A \subset Var \subseteq F(\L)$ are the constant propositional symbols  as well, and we will use the same symbols for them as those
used for elements in $A$, with the bottom and top elements $0,1$
respectively. Lindenbaum algebra of $\L$ is the quotient algebra
$F(\L)/\equiv$, where for any two formulae $\phi, \psi \in F(\L)$,
it holds that $~~\phi \equiv \psi~~$ iff $~~\phi \Vvdash \psi$ and
$\psi \Vvdash \phi~$.\\
The   algebraic existential \emph{modal}
operators $o_i:A \rightarrow A, i = 1,2,...,$ are monotonic,
additive ($o_i(x \vee y) = o_i(x) \vee o_i(y)$)
 and $o_i(0) = 0$) (the universal modal operators are monotonic and  multiplicative $\overline{o}_i(x \wedge y) = \overline{o}_i(x) \wedge \overline{o}_i(y)$,
  $\overline{o}_i(1) = 1$).
  They appear often in many-valued logics, for example,
  as conflation operator (knowledge negation \cite{Fitt91})  and Moore's autoepistemic operator \cite{Gins90} in Belnap's 4-valued bilattice \cite{Beln77} (in Example 8),
  or modal operators $L$ and $M$ of Lukasiewicz's 4-valued logic \cite{Luka53,Luka57,MaPr09}, or recently in \cite{GeJo94,Dunn95,DuZh05,Rest05}. The correspondent set of \emph{logical}
modal operators (existential and universal) will be denoted in
  standard way by $\lozenge_i \in \O$ and $\square_i \in \O$ respectively.\\
 A many-valued \emph{modal} logic here is a truth-functional many-valued logic with
 a non-empty set of  modal operators with properties defined
 above. \\
 A  \emph{valuation} $I$ as a mapping $I: Var
\rightarrow A$  such that for any $x \in A$, $\overline{I}(x) = x$.
It can be uniquely extended to the homomorphism $\overline{I}: F(\L)
\rightarrow A$ (for example,
 for any $p,q \in Var$, $\overline{I}(p \odot q) = I(p) \odot I(q), \odot \in \{\wedge, \vee, \Rightarrow \}$,
 $\overline{I}(\neg p) = \neg I(p)$,  and $\overline{I}(\lozenge_i p) = o_i( I(p))$, where
 $\wedge, \vee, \Rightarrow, \neg$ are conjunction, disjunction, implication and negation respectively). We denote by $\mathbb{V}_m$ the set of all   valuations in $A^{Var}$.\\
\textbf{Relevant work}: we will briefly present the previous work,
based on
algebraic \emph{matrices}, and explain some weak points of such a matrix-based approach.\\
  The standard approach to representation
theorems uses a subset $D \subset A$ of the set of truth values $A$,
denominated designated elements; informally the
designated elements represent the equivalence class of the theorems
of $\L$. Given an algebra $\textbf{A} = (A, \{o\}_{o \in \O})$, the
$\O$-matrix is the pair $(\textbf{A},D)$, where $D \subset A$ is a
subset of designated elements. The \emph{algebraic} satisfaction
relation $\models^a$ ('a' stands for 'algebraic') is defined as
follows:
\begin{definition} \label{def:matr} Let $\L = (Var, \O,
\Vvdash)$ be a logic, $(\textbf{A},D)$ a $\O$-matrix, and $\phi \in
F(\L)$. Let $I:Var \rightarrow A$ be a map that assigns logic values
to propositional variables, and $\overline{I}:F(\L) \rightarrow A$
be its unique extension to all formulae in a language $\L$. Let $\M$
be a class of $\O$-matrices. We  define  the relation $\models^a$
inductively as
follows:\\
1. $(\textbf{A},D)\models^a_I\phi~$ iff $~ \overline{I}(\phi)
\in D$,\\
2. $(\textbf{A},D)\models^a\phi~$ iff $~\overline{I}(\phi)
\in D~$ for every $~I:Var \rightarrow A$,\\
3. $\M \models^a\phi~$ iff $~(\textbf{A},D)\models^a~\phi~$
for every $~(\textbf{A},D) \in \M$.\\
A logic $\L$ is sound w.r.t. $\M$ iff for every $~\phi \in F(\L)$,
if $~\L \Vvdash \phi$ then $~\M \models^a\phi$.\\
$\L$ is complete w.r.t. $\M$ iff for every  $~\phi \in F(\L)$, if
$~\M \models^a\phi$ then $~\L \Vvdash \phi$.
\end{definition}
 Dual to
algebraic semantics, based on the class $\M$ of $\O$-matrices we
also have the Kripke-style semantics based on a class $\R$ of
relational models where the satisfiability relation $~\models^r~$ is
defined by induction on the structure of the formulae.
Substantially, each relational model $K \in \R$ is a Kripke frame
over a set of possible worlds with additional accessibility
\emph{relations} between possible worlds associated with logical
operators. The distinctive feature of this relational semantics is
that the accessibility relations are used in the definition of
satisfiability, which is not just a mechanical truth-functional
translation of the formula structure into the
model.\\
 The definition of the algebraic/relational duality  is based on the following assumption:
\begin{definition} \textsc{Representation Assumption \cite{Sofr03}} \label{def:abs}\\
Assume that there exists a class $\R$ of relational structures such
that there exist $\mathbb{D}:\M \rightarrow\R$, $\mathbb{E}:\R
\rightarrow \M$ such that (C):\\
(i)  for every $K \in \R$, $\mathbb{E}(K) = (\textbf{A}_K, D_K) \in
\M$, where $\textbf{A}_K$ is an algebra of subsets of the support of
$K$;\\
(ii) for every $M = (\textbf{A},D) \in \M$, if
$\mathbb{E}(\mathbb{D}(M)) = (\textbf{A}_{\mathbb{D}(M)},
D_{\mathbb{D}(M)})$ then there is an injective homomorphism
$~i_n:\textbf{A} \rightarrow \textbf{A}_{\mathbb{D}(M)}$ with
$~i_n^{-1}(D_{\mathbb{D}(M)}) \subseteq D$.
\end{definition}
Let $m:Var \rightarrow A_K$ be a meaning function (assigns logic
values to propositional variables), then $~(K,m)~$ is the Kripke
model for a frame $K$. Then, the definition of the relation
$\models^r$ can be given as follows:
\begin{definition} \cite{Sofr03} Assume that $\M$ and $\R$ satisfy
condition (C)(i). Let $\K \in \R$,  $m:Var \rightarrow A_K$, and
$~\overline{m}:F(\L) \rightarrow A_K$ be the unique homomorphism of
$\O$-algebras that extends $~m$. Let $y$ be an element in the
support of $K$, Then:\\
1. $~~~K \models^r_{m,y}~\phi~~~$ iff  $~~~y \in
\overline{m}(\phi)$;\\
2. $~~~K \models^r_{m}~\phi~~~$ iff  $~~~\overline{m}(\phi) \in
D_K$;\\
3. $~~~K \models^r~\phi~~~$ iff  $~~~$ for every
$m$, $~K \models^r_{m}~\phi~$.
\end{definition}
A logic $\L$ is sound w.r.t. $\R$ iff for every $~\phi \in F(\L)$,
if $~\L \Vvdash \phi$ then $~\R \models^r~\phi$. $~\L$ is complete
w.r.t. $\R$ iff for every  $~\phi \in F(\L)$, if $~\R
\models^r~\phi$ then $~\L \Vvdash \phi$.\\
In \cite{Sofr03} it is demonstrated that if $\L = (Var, \O,
\Vvdash)$ is sound and complete w.r.t. a class $\M$ of
$\O$-matrices, and there exists a class $\R$ such that the
Assumption (C) holds, then $\L$ is sound and complete w.r.t. the
class of Kripke-style models $\K_{\M, \R} = \{(K,m)|~K \in \R, m:Var
\rightarrow A_K$
 $~$ where $ ~~\mathbb{E}(K) = (\textbf{A}_K,D_K) \}$.
\\The strong and weak points of this approach:
\begin{itemize}
  \item In a matrix-based many-valued logic, a formula is \emph{satisfied} if its logic value is a designated
  value.
  Such an approach, \emph{based on} $\O$\emph{-matrices}, is very effective for all kinds of 2-valued logic
  where the set of designated elements is a singleton set composed by only true value, $D =
  \{1\}$, as in the case of classical, intuitionistic and 2-value
  modal logics (extension of Boolean algebra). It is only a partially
  good  solution for the case when a set of truth-values \emph{can not} be
  easily divided into two complementary subsets: $D \subset A$ for
  values for which we retain that a formula can be considered
  satisfied, and its complement $A\backslash D$ for those which we retain that a formula cannot be considered satisfied. For example, in the case
  of fuzzy logic where $A = [0,1]$ (the closed set of reals between 0 and
  1)  we can assume that $D = [a, 1]$ is the closed set between
  some prefixed value $0 < a < 1$ and 1. But is not clear what is the \emph{correct} value for
  $a$ for generally acceptable fuzzy logic
  (otherwise we will have an infinite number of different logics for each an arbitrary value $a$).\\
   An analog difficulty we can find in the case of bilattices  \cite{Gins88,Fitt91,ArAv94,RuFa94,LaSa94,KSim01,Majk05Fu}.
\item  The
  second observation  is that the
  representation theorems define the
  isomorphism between a many-valued algebra and the
  set-based algebra that is a subalgebra of the canonical extension
  of the original many-valued algebra. It will be useful
  to define directly such an isomorphism based on the
  duality assumption (C).
\end{itemize}
 \textbf{Main contribution}: The main contribution, presented in
 Section 3, is a general representation theorem for many-valued logics
 with the \emph{truth-invariance} entailment for \emph{any} set of truth-values $A$ (also if it is not a lattice).
  It is \emph{substantially} different w.r.t the previous representation theorems that are all based on
 matrices, and is based on algebraic models of a logic. We
replace the duality Algebras (Matrices) - Relational structures
described in previous work, by the semantic duality Algebraic models
- Kripke models. The novelty is that the set of models of a given
logic can be obtained by using Gentzen-like sequent calculi
\cite{Majk09BS} without  using necessarily the subset of designated
elements (matrices). As a guiding
 \emph{example} instead, here in Section 2, will be presented a more specific
 case, when a  logic is based on complete distributive lattice $A$.
 These sequent-based
representations of many-valued logics with truth-invariance
entailment allows us to define, without using the matrices, the set
of \emph{models} of a given many-valued logic, required by  general
definitions in Section 3. This particular example in Section 2, when
$A$ is complete distributive lattice, is then used in Section 4 for
a concrete definition of Kripke frames based on an autoreferential
assumption \cite{Majk06ml} where the set of possible worlds is fixed
by a subset of algebraic truth values in $A$.
\\
This paper is based on the idea that the satisfaction relation (and
the entailment) in the case of many-valued logics can be defined
\emph{without} using the subset of designated elements.
 For example, in the case of logic programs,
let $I:Var \rightarrow A$ be a many-valued valuation, and $(A,
\sqsubseteq)$ be the set $A$ of logic values with
 partial truth order $\sqsubseteq$. Then, given any rule $B
\leftarrow B_1 \wedge ... \wedge B_n$ where $B$ is a propositional
letter and $B_i$ is a ground literal (propositional letter or
negation of them), we say that it is \emph{satisfied}  iff $I(B)
\sqsupseteq I(B_1) \wedge ... \wedge I(B_n)$; the valuation that
satisfies all rules is a \emph{model} for such a logic program. As
we have seen in this case, instead of the subset $D \subseteq A$ of
designated
elements, we simply use the truth ordering between logic values.\\
  The simple way to extend this example to any
propositional logic $\L = (Var, \O, \Vdash)$ is to consider
equivalently this logic as a sequent system of (structural and
logical) rules $R: \frac{s_1, ..., s_k}{s}$ where each $s_i$ is a
sequent $\phi_1,..,\phi_n \vdash \psi_1,...,\psi_m$ where,
accordingly to Gentzen, the commas in the left are conjunctions
while those on the right are disjunctions, and $\phi_i, \psi_j \in
F(\L)$ are logic formulae. We say that a valuation $I:Var
\rightarrow A$ \emph{satisfies} this sequent iff
$\overline{I}(\phi_1) \wedge..\wedge \overline{I}(\phi_n)
\sqsubseteq \overline{I}(\psi_1) \vee...\vee \overline{I}(\psi_m)$,
and that $I$ satisfies a rule $R$ iff $I$ satisfies the conclusion
sequent $s$ of this rule whenever it satisfies all sequent premises
$s_1, ..., s_k$ of this rule. Then, a \emph{model} of this logic is
any valuation $I$ which satisfies all
 logic sequent rules of this logic (the structural sequent rules as Identity, Cut, Weakening, Permutation, Contraction and Associativity
 rules are satisfied by all valuations).\\
Notice that this sequent-based approach is always possible,
independently of the algebraic properties of the set of truth-values
in $A$, for example by transforming the original many-valued logic
into 2-valued modal logic \cite{MajkC04,Majk06MV}, and defining the
classical 2-valued sequent rules as presented in \cite{Majk09BS}
with the truth-invariance entailment for many-valued logics. This
\emph{truth-invariance entailment} will be used for a new
representation theorem in this paper (in Definition
\ref{def:newalg}). Notice that the sequent system can be used also
as a basis for an \emph{autoreferential} algebraic/relational
semantics of many-valued logics \cite{Majk06ml}.\\
In what follows we denote by $y \sqsubset x$ iff ($y \sqsubseteq x$
and not $x \sqsubseteq y$), and we denote by $x \bowtie y$ two
unrelated elements in $A$ (so that not $(x\sqsubseteq y$ or
$y\sqsubseteq x)$).\\
We define the following mapping $\downarrow:A \rightarrow \P(A)$
such that for any  element $x \in A$, we obtain the closed set
$\downarrow x = \{a \in A~| a \sqsubseteq x~\}$. It is well known
that for any two elements of a complete lattice $x, y \in A$ holds
the set intersection closure property $ \downarrow x \bigcap
\downarrow y = \downarrow (x \wedge y)$, but \emph{does not} hold
the union closure property, that is, generally  does not exists $z
\in A$  such
that $ \downarrow x \bigcup  \downarrow y = \downarrow z$. \\
But the closure property for the intersection and union holds for
the more general case of \emph{hereditary subsets}: a set $B \in
\P(A)$ is hereditary if it is closed downwards under $\sqsubseteq$,
i.e., if we have that whenever $x \in B$ and $y \sqsubseteq x$ then
$y \in B$. Notice that the bottom hereditary subset of any complete
lattice $A$ is the set $\downarrow 0 = \{ 0 \}$ where $0$ is the
bottom element of A. Thus while $\P(A)$ is a topological space, its
subset composed by only hereditary subsets of $A$, used to define
the canonical representation isomorphic to the algebra $\textbf{A}$,
there will not be topological space (because the empty set will not
be an element of the carrier set of this canonical subalgebra of the
powerset canonical extension
algebra). This is also seen in power-domains in the domain theory, where the empty set is often excluded.\\
This paper follows the following plan:\\In Section 2 we show, in a
particular example, how we are able to avoid the matrices used
 in previous Representation theorem frameworks for many-valued logics: we present an autoreferential semantics for many-valued logics, based on
 sequents and many-valued \emph{valuations}. In Section 3 we
define the main result of this paper: a new general
 Representation theorem framework for many-valued logics with truth-invariance entailment, where we replace the duality
Algebras (Matrices) - Relational structures by the semantic duality
Algebraic models - Kripke models.
 In Section 4 we  apply this new Representation theorem framework  to
 modal many-valued logics, in the particular case when it is based on complete distributive lattices of truth
 values (an autoreferential representation). In Section 5 we consider a concrete example
 of Belnap's bilattice, composed by two (truth and knowledge) complete distributive lattices, used for  applications in logic programming
 with incomplete and inconsistent information.
%
\section{Sequents for Many-valued logics based on complete distributive lattices}
The main result of this work is a new representation theorem for
\emph{any} many-valued logic, based on \emph{models} of such a
logic, and will be presented in the next section. In this section
instead we will introduce an example of defining the set of models
of a given many-valued logic $\L$, based on binary sequent systems
for many-valued logics. \\
Sequent calculus, introduced by Gentzen \cite{Gent32} and Hertz
\cite{Hert29} for classical logic, was generalized to the
many-valued (m-sequents) case by Rouseau \cite{Rous67} and others.
The tableaux calculi were presented in \cite{Carni87,Hanh91}. The
strict correspondence between the cut-free m-sequent calculus and
closed tableaux has been presented in \cite{BFZa93a}. The more
detailed information for interested readers can be found in
\cite{BFZa93,BFSa00}. This ad-hoc m-sequent system is not standard
one. Consequently, it is interesting to consider a calculus for
many-valued logics based on standard \emph{binary} sequents.
   Such a standard two-sides sequent
  calculi for  lattice-based many-valued logics
    has been  elaborated  recently (with an autoreferential Kripke-style semantics for such
  logics) in two complementary ways in
  \cite{Majk06ml,Majk08dC}.\\
A sequent system for the \emph{truth-invariance} semantics of the
entailment, used in a new representation theorem in the next
Section, was recently presented in \cite{Majk09BS}. Such a general system does not use the partial ordering of the truth values in $A$. \\
 Here we will present another example of a sequence system, for many-valued logics with a
complete distributive lattice $A$, with truth-preserving semantics
of the entailment. It is another example, more specific than that in
\cite{Majk09BS},  of how we can define
the \emph{models} of many-valued logics without using the matrices.\\
We justify this significant case because the algebras for all
many-valued logics with finite set of logic values are complete
lattices. And also the algebras for fuzzy logic, belief based logic,
etc., \cite{Majk06Bi}, with \emph{infinite} number of logic values,
are complete and distributive lattices as well.
In what follows we will use the approach in \cite{Majk08dC}, with
the valuation-based semantics for many-valued logic.
 Given a propositional logic $\L$  a
sequent
 is a consequence pair of formulae $ s = (\phi ; \psi) \in F(\L) \times F(\L)$, denoted also by $\phi \vdash
 \psi$. \\A Gentzen system, denoted by a pair $\G = \langle \mathbb{L}, \Vdash
 \rangle$, where $\Vdash$ is finitary consequence relation on set of sequents in $\mathbb{L} = F(\L) \times F(\L)$,
 is said to be \emph{normal} if it satisfies the following
 conditions: for any sequent $s = \phi \vdash
 \psi \in \mathbb{L}$ and a set of
 sequents $\Gamma \subseteq \mathbb{L}$, \\
 1. (reflexivity) if $s \in \Gamma$ then $\Gamma \Vdash s$\\
 2. (transitivity) if $\Gamma \Vdash s$ and for every $s' \in
 \Gamma$, $\Theta \Vdash s'$, then $\Theta \Vdash s$\\
 3. (finiteness) if $\Gamma \Vdash s$ then there is finite $\Theta
 \subseteq\Gamma$ such that $\Theta \Vdash s$.\\
 4. for any homomorphism $\sigma$ from $\mathbb{L}$ into itself
 (i.e., substitution), if $\Gamma \Vdash s$ then $\sigma[\Gamma] \Vdash
 \sigma(s)$, i.e., $\{\sigma(\phi_i) \vdash \sigma(\psi_i)~|~\phi_i \vdash
 \psi_i \in \mathbb{L}\} \Vdash (\sigma(\phi) \vdash \sigma(\psi))$.\\
 Notice that from (1) and (2) we obtain the monotonic property:\\
 5.  if  $\Gamma \Vdash s$ and $\Gamma
 \subseteq\Theta$, then $\Theta \Vdash s$.\\
 We denote by $C:\P(\mathbb{L})\rightarrow \P(\mathbb{L})$ the
 closure operator such that $C(\Gamma) =_{def} \{ s \in \mathbb{L}~|~\Gamma \Vdash
 s\}$, with the properties: $\Gamma \subseteq C(\Gamma)$ (from
 reflexivity (1)); it is monotonic, i.e., $\Gamma \subseteq \Gamma_1$ implies
 $C(\Gamma) \subseteq C(\Gamma_1)$ (from  (5)), and  $C(C(\Gamma)) = C(\Gamma)$ as well. Thus, we obtain \\
 6. $\Gamma \Vdash s ~$ iff $~s \in C(\Gamma)$.\\
 Any sequent theory $\Gamma \subseteq \mathbb{L}$ is said to be a \emph{closed}
 theory iff $\Gamma = C(\Gamma)$. This closure property corresponds to the
 fact that $\Gamma \Vdash s$ iff $s \in \Gamma$.\\
 Each sequent theory $\Gamma$ can be considered as a bivaluation (characteristic function) $\beta:\mathbb{L} \rightarrow
 \textbf{2}$ such that for any sequent $s \in \mathbb{L}$, $~\beta(s)
 = 1~$ iff $~s \in \Gamma$.
We will use this 2-valued valuation-based semantics in order to
define the sound and complete \emph{many-valued} valuation-based
semantics for many-valued modal logics.\\
 \textbf{Example 1:} Let us consider the many-valued modal logic with a distributive complete lattice $(A,\sqsubseteq)$ of
  truth values (where all truth-values in $A$ are language primitives as well), that is an
 extension of the Distributive modal logic (distributive lattice
 logic DDL)  \cite{Dunn95,GNVe05}
 (with $\Box$ universal modal operator, and its left adjoint
 existential modal operator $\Diamond$, with $\Diamond \dashv \Box$)
 and with \emph{negative} modal additive operator $\neg:(A,\sqsubseteq,\vee)\rightarrow(A,\sqsubseteq,\vee)^{OP}$, where $\vee^{OP} = \wedge$, $\sqsubseteq^{OP} = \sqsupseteq$. The binary consequence
 system $\G$,  in this logic  $\L$, is as follows:\\
 (AXIOMS) $\G$ contains  the following sequents:\\
 1. $\phi \vdash \phi~~$ (reflexive)\\
 2. $\phi \vdash 1$, $~~0 \vdash \phi~~$ (top/bottom axioms)\\
  3. $\phi \wedge \psi \vdash \phi$, $~~\phi \wedge \psi \vdash \psi~~$ (projections: axioms for
 meet)\\
 4. $\phi  \vdash \phi \vee \psi$, $~~\psi \vdash \phi \vee \psi~~$ (injections: axioms
 for join)\\
 5. $\phi \wedge (\psi \vee \varphi) \vdash (\phi \wedge \psi) \vee (\phi \wedge \varphi)~~$
 (distributivity axiom)\\
 6. $\Box (\phi \wedge \psi) \vdash \Box \phi \wedge \Box \psi$, $~~1 \vdash \Box
 1~~$ (multiplicative modal property axioms)\\
 7.  $\Diamond (\phi \vee \psi) \vdash \Diamond \phi \vee \Diamond \psi$, $~~\Diamond 0  \vdash
 0~~$ (additive modal property axioms)\\
 8. $\neg \phi \wedge \neg \psi \vdash \neg (\phi \vee \psi)$, $~~1 \vdash \neg 0$
 (additive modal negation axiom)\\
  9. The set of sequents that define the poset of the lattice of
 truth values $(A, \sqsubseteq)$: for any two $x,y \in A$, if $x \sqsubseteq y$
 then $x \vdash y$ is an axiom.\\\\
 (INFERENCE RULES) $\G$ is closed under the following inference
 rules:\\
 1. $~~\frac{\phi ~\vdash \psi,~~\psi~ \vdash \varphi }{\phi ~\vdash \varphi}~~$ (cut/
 transitivity rule)\\
 2. $~~\frac{\phi ~\vdash \psi,~~\phi~ \vdash \varphi}{\phi~ \vdash \psi \wedge
 \varphi}~~$, $~~\frac{\phi ~\vdash \psi,~~\varphi~ \vdash \psi}{\phi \vee \varphi~ \vdash \psi
 }~~$ (lower/upper lattice bound rules)\\
 3. $~~\frac{\phi ~\vdash \psi}{\Box \phi ~\vdash \Box \psi}~~$, $~~\frac{\phi ~\vdash
 \psi} {\Diamond \phi ~\vdash \Diamond \psi }~~$ (monotonicity of
 modal operators rules)\\
 4. $~~\frac{\phi ~\vdash \psi}{\neg \psi ~\vdash \neg \phi}~~$ (antitonicity of
 modal negation rule)\\
 5. $~~\frac{\phi~\vdash \psi}{\sigma(\phi) ~ \vdash \sigma(\psi)} ~~$ (substitution
 rule: $\sigma$ is substitution $(\gamma/p)$).\\\\
 Notice that the rules in point 2 are the consequences of the diagonal mapping $\triangle: A \rightarrow Y$, where $Y = A\times A$ and $\triangle x = (x,x)$,
  (which is both an additive and multiplicative modal operator), and its Galois
 adjunctions with the meet (multiplicative) and join (additive) operators $\wedge, \vee: Y\rightarrow A$, i.e., with $\triangle\dashv
 \wedge$ and $\vee\dashv \triangle$; that is $\triangle x \leq_Y
 (y,z)$ (i.e., $x \sqsubseteq y$ and $x \sqsubseteq z$) iff $x \sqsubseteq \wedge (y,z) = y \wedge z$,
 and $x \vee y = \vee(x,y) \sqsubseteq z$ iff $(x,y) \leq_Y \triangle z$ (i.e., $x
 \sqsubseteq z$ and $y \sqsubseteq z$).
 \\$\square$ \\
 The axioms from 1 to 5 and the rules  1 and 2 are
 taken from \cite{Dunn95} for the $DLL$ and it was shown that this sequent-based Gentzen-like system is sound and
 complete. The system $\G$ in Example 1 in only a guiding example, that will be consider in the rest of this section. We are able  to introduce another logical
 connectives for any given many-valued modal logic (where existential modal operators are monotonic, additive and normal) based on the
 complete distributive  lattice of truth values in $A$, in order to obtain a
 similar sequent system as this in Example 1.
 Notice that  in a  Gentzen-like deductive system $\G$ above each
 sequent is a valid truth-preserving consequence pair defined by the poset
 of the complete distributive lattice $(A, \sqsubseteq)$ of truth values (which are also the
 constants of this modal propositional language $\L$). Consequently, each
 occurrence of the symbol $\vdash$ can be substituted by the partial
 order $\sqsubseteq$ of this complete lattice.
 \begin{definition} \textsc{Truth-preserving entailment:}
 \label{def:entailment} For any two formulae $\phi, \psi \in F(\L)$,
 the truth-preserving consequence pair (sequent) denoted by
 $\phi \vdash \psi$ is satisfied by a given valuation $I \in \mathbb{V}_m$ if $~~\overline{I}(\phi) \sqsubseteq \overline{I}(\psi)$.\\
  This sequent is a
 tautology if it is satisfied by all valuations, i.e., when
 $~\forall I \in \mathbb{V}_m (\overline{I}(\phi) \sqsubseteq \overline{I}(\psi))$.\\
 For a  normal Gentzen-like sequent system $\G $ of the many-valued logic $\L$, with the set of sequents $Seq_{\G} \subseteq
 \mathbb{L}$ and a set of inference rules in $Rul_{ \G}$, we say that a many-valued valuation $I$ is its \verb"model"
  if it satisfies all sequents in $\G$. The set of all models of
  a given set of sequents (theory) $\Gamma$ is denoted by
  $~~Mod_{\Gamma}  = \{I \in \mathbb{V}_m|\forall(\phi \vdash \psi) \in \Gamma(\overline{I}(\phi) \sqsubseteq \overline{I}(\psi))
  \} \subseteq \mathbb{V}_m \subset A^{Var}$.
 \end{definition}
  \begin{propo} \textsc{Soundness:} \label{prop:sound}
 All  axioms of the Gentzen-like sequent system $\G $  of a many-valued logic $\L$ based on complete distributive lattice
 $(A, \sqsubseteq)$ of algebraic truth values are the tautologies, and all its rules are sound for the model satisfiability and
 preserve the tautologies.
 \end{propo}
 \textbf{Proof:} It is straightforward to verify (see the Example 1) that all axioms are tautologies (all constant sequents
  specify the poset of a complete lattice $(A, \sqsubseteq)$, thus are tautologies). It is straightforward to verify that all rules preserve the tautologies.
  Moreover, if all premisses  of any rule in $\G$ are satisfied by a given many-valued valuation $\overline{I}:F(\L)\rightarrow A$,
  then also the deduced sequent of this rule is satisfied by the same valuation, i.e., the rules are sound for the model satisfiability.
  \\$\square$ \\
 It is easy to verify that for any two $x,y \in A$ we have that $x
 \sqsubseteq y$ iff $x \vdash y$, that is the truth-preserving entailment
 coincides with the partial truth-ordering in a lattice $(A, \sqsubseteq)$.
 Notice that
 it is compatible with the lattice operators, that is, for any two formulae $\phi, \psi \in \L$,
 $~\phi \wedge \psi \vdash \psi$ and $~\phi \vdash \psi \vee \phi$.
 This entailment imposes the following restrictions on the logic implication: in order to satisfy the Deduction Theorem
 "$ ~z \vdash x \Rightarrow y~$ iff $~z \wedge x \vdash y$"
 (i.e., inference rules for elimination and introduction of the logic connective $\Rightarrow$,
  $~~\frac{z ~\vdash x \Rightarrow y}{z \wedge x ~\vdash y}~~$ and $~~\frac{z \wedge x ~\vdash y}{z ~\vdash x \Rightarrow y}~$)
  by this entailment,
 the logic implication must satisfy (the case when $z = 1$) the requirement that for any
 $x,y \in X$, $~x \Rightarrow y = 1~$ iff $~ x \sqsubseteq y$, while
 it must satisfy $x \wedge (x \Rightarrow y) \sqsubseteq y$ in order to satisfy the Modus Ponens inference rule.
 The particularity of this entailment is that any consequence pair (sequent) $\phi \vdash
 \psi$ is algebraically an equation $\phi \wedge \psi = \phi$ (or, $\phi \vee \psi = \psi$).
  It is easy to verify, that in the case of the classical 2-valued
 propositional logic this entailment is equal to the classical
 propositional entailment. Consequently, this truth-preserving entailment is
 only a generalization of the classical entailment for a many-valued
 propositional logics.\\
  \textbf{Remark:} It is easy to observe that each sequent is, from
 the logic point of view, a  \emph{2-valued object} so that all inference
 rules are embedded into the classical 2-valued framework, i.e., given a bivaluation $\beta:\mathbb{L} \rightarrow \textbf{2}$,
 we have that a sequent $s = \phi \vdash \psi$ is satisfied when $\beta(s) = 1$, so that we have the relationship between sequent
  bivaluations and many-valued
 valuations $I$ used in Definition \ref{def:entailment}.
 In fact we have that $\beta = eq \circ
<\pi_1, \wedge >\circ (\overline{I} \times \overline{I}):\mathbb{L}
\rightarrow \textbf{2}$ is the characteristic function, with $\pi_1$
first projection, a valuation $\overline{I}:F(\L)\rightarrow A$, and
$eq:A \times A \rightarrow \textbf{2}$ (defined by $eq(a,b) = 1$ iff $a = b$).\\
Consequently, $\beta(\phi \vdash \psi) = \beta (\phi ; \psi) = eq
\circ <\pi_1, \wedge
>\circ(\overline{I} \times \overline{I})(\phi ; \psi) = eq \circ <\pi_1, \wedge >(\overline{I}(\phi)
, \overline{I}(\psi)) = \\
= eq <\pi_1(\overline{I}(\phi) , \overline{I}(\psi)),
\wedge(\overline{I}(\phi) , \overline{I}(\psi))
> =\\ = eq(\overline{I}(\phi), \wedge(\overline{I}(\phi) , \overline{I}(\psi))) = eq(\overline{I}(\phi), \overline{I}(\phi)
\wedge \overline{I}(\psi))$.\\
Thus $\beta(\phi \vdash \psi) = 1 $ iff $\overline{I}(\phi)
\sqsubseteq \overline{I}(\psi)$, i.e., when this sequent is satisfied by $I$.\\
  From my
 point of view, this sequent feature, which is only an alternative formulation
 for the 2-valued classical logic, is \emph{fundamental} in the framework of
 many-valued logics, where  the semantics for the entailment,
 based on algebraic
 matrices $(\textbf{A},D)$ is often arbitrary.
 \\$\square$\\
 Thus, this correct definition of the 2-valued entailment in the sequent system $\G$, based only on the lattice
 ordering, can replace the current entailment based on the algebraic
 matrices $(\textbf{A},D)$, where $D \subseteq A$ is a subset of designated
 elements, which is upward closed. That is, if $x \in D$ and $x \sqsubseteq
 y$ then $y \in D$ (thus $ 1 \in D$). Consequently, the matrix-entailment,  defined by $\phi \vdash_D \psi$,
 is valid iff $\forall I \in \mathbb{V}_m.(\overline{I}(\phi) \in D$ implies $ \overline{I}(\psi) \in D)$.
 It is easy to verify also that $\phi \vdash \psi$ implies $\phi \vdash_D \psi$.
 Thus, we are now able  to introduce the model-theoretic semantics for the many-valued logics:
\begin{definition}   \label{def:manyvaluation}
 A \verb"many-valued"  model-theoretic semantics of a given many-valued logic $\L$, with a Gentzen system $\G = \langle \mathbb{L}, \Vdash
 \rangle$,
 is a semantic deducibility relation
 $~\models_m$,
 defined for any $\Gamma \subseteq \mathbb{L}$ and sequent $~s = (\phi \vdash \psi) \in \mathbb{L} ~~$
 by:\\ $~\Gamma \models_m s~~$ iff $~~~~$"all many-valued models of $\Gamma$ are the models of
 $s$"\\
 iff $~~~~\forall I \in
 \mathbb{V}_m (~ \forall(\phi_i \vdash \psi_i) \in \Gamma(\overline{I}(\phi_i) \sqsubseteq
 \overline{I}(\psi_i))~$ implies $~\overline{I}(\phi) \sqsubseteq \overline{I}(\psi))~~~~$
 iff $~~~~\forall I \in
 Mod_{\Gamma} (~ \forall (\phi_i \vdash \psi_i) \in \Gamma(\overline{I}(\phi_i) \sqsubseteq
 \overline{I}(\psi_i))~$ implies $~\overline{I}(\phi) \sqsubseteq \overline{I}(\psi))$\\
 iff $~~~~\forall I \in
 Mod_{\Gamma} (~\overline{I}(\phi) \sqsubseteq \overline{I}(\psi))$.
\end{definition}
\textbf{Example 2:} Let us consider a many-valued logic with
$\{p,q,r,r_1\} \subseteq Var$  and many-valued clauses $P_r = \{p
\Leftarrow a, q
 \Leftarrow b, r \Leftarrow p, r\Leftarrow q \}$, with $a, b \in A$. The sequent-based
 translation of $P_r$ results in a sequent theory $\Gamma = \{p
 \vdash a, a \vdash p, q \vdash b, b \vdash q, r \vdash p \vee q, p
 \vee q \vdash r \}$, so that the set of models of $P_r$ is equal to
 $~~Mod_{\Gamma} = \{I:Var \rightarrow A ~|~I(p) = a, I(q) = b, I(r) = a \vee b$ and $I(r_1) \in A
 \}$. Thus we have that $~\Gamma \models_m (p \wedge q \vdash a \wedge
 b)~$ and $~\Gamma \models_m (a \wedge b \vdash p \wedge
 q)~$, while for every $c \in A$, $~\Gamma \nvDash_m (r_1 \vdash
 c)$.
 \\$\square$

It is easy to verify that the Gentzen-like system $\G = \langle
\mathbb{L}, \Vdash  \rangle$ of a complete-lattice based many-valued
 is a \emph{normal} logic.
\begin{theo}   \label{prop:manyvaluation}
The many-valued model theoretic semantics is an adequate semantics
for a many-valued logic $\L$ specified by a Gentzen-like logic
system $\G = \langle \mathbb{L}, \Vdash
 \rangle$, that is, it is sound and complete. \\Consequently,
$~~~\Gamma \models_m s~~$ iff $~~\Gamma \Vdash s$.
\end{theo}
\textbf{Proof:} Let us prove that for any many valued model $I \in
Mod_{\Gamma}$, the obtained sequent bivaluation $\beta = eq \circ
<\pi_1, \wedge >\circ (\overline{I} \times
\overline{I}):\mathbb{L}\rightarrow \textbf{2}$ is the
characteristic function of the closed theory $\Gamma_{I} = C(T)$
with $T =  \{ \phi \vdash x, ~x \vdash \phi~|~\phi \in \L,~ x =
\overline{I}(\phi) \}$. From the definition of $\beta$ we have that
$\beta(\phi \vdash \psi) = \beta (\phi ; \psi) = eq \circ <\pi_1,
\wedge
>\circ(\overline{I} \times \overline{I})(\phi ; \psi) = eq \circ <\pi_1, \wedge >(\overline{I}(\phi)
, \overline{I}(\psi)) = eq <\pi_1(\overline{I}(\phi) ,
\overline{I}(\psi)), \wedge(\overline{I}(\phi) , \overline{I}(\psi))
> = eq(\overline{I}(\phi), \wedge(\overline{I}(\phi) , \overline{I}(\psi))) = eq(\overline{I}(\phi), \overline{I}(\phi)
\wedge \overline{I}(\psi))$.\\
Thus $\beta(\phi \vdash \psi) = 1 $ iff $\overline{I}(\phi) \sqsubseteq \overline{I}(\psi)$, i.e., when this sequent is satisfied by $I$.\\
1. Let us show that for any sequent $s$, $~~ s\in \Gamma_I$ implies
$~\beta(s) = 1$: First of all any sequent $s \in T$ is of the form
$\varphi \vdash x$ or $x \vdash \varphi$, where $x = I(\varphi)$, so
that it is satisfied by $I$ (it holds that $\overline{I}(\varphi)
\sqsubseteq \overline{I}(\varphi)$ in both cases). Consequently,
\emph{all} sequents in $T$ are satisfied by $I$. From Proposition
\ref{prop:sound} we have that all inference rules in $\G$ are sound
w.r.t. the model satisfiability, thus for any deduction $T \Vdash s$
(i.e., $s \in \Gamma_I$) where all sequents in premisses are
satisfied by the many-valued valuation (model) $I$, also the deduced
sequent $s = (\phi \vdash \psi)$
must be satisfied, that is, it must hold that $\overline{I}(\phi) \sqsubseteq \overline{I}(\psi)$, i.e., $\beta(s) = 1$.\\
2. Let us show that for any sequent $s$,  $~\beta(s) = 1$ implies
$~~ s\in \Gamma_I$:
 For \emph{any} sequent $s =(\phi \vdash \psi) \in \mathbb{L}$ if $~\beta(s) = 1$ then $x = \overline{I}(\phi) \sqsubseteq \overline{I}(\psi) = y$
 (i.e., $s$ is satisfied by $I$).
From the definition of $T$, we have that $\phi \vdash x, y \vdash
\psi \in T$, and from $ x\sqsubseteq y$ we have  $x \vdash y \in
Ax_{\G}$ (where $Ax_{\G}$ are axioms (sequents) in $\G$, with $\{ x
\vdash y~|~x,y \in A, x \sqsubseteq y \} \subseteq Ax_{\G}$,
 thus satisfied by every valuation) by the transitivity rule we obtain that $T \Vdash (\phi \vdash
\psi)$,
 i.e., $s =(\phi \vdash \psi) \in  C(T) = \Gamma_I$.
So, from (1) and (2) we obtain that $~~~\beta(s) = 1~~$ iff $~~ s\in
\Gamma_I$, i.e., the sequent bivaluation $\beta$
 is the characteristic function of a closed set.
  Consequently, any many-valued \emph{model} $v$ of this many-valued logic $\L$
corresponds to the \emph{closed} bivaluation $\beta$ which is a
characteristic function of a closed theory of sequents:  we define
the set of all closed bivaluations obtained from the set of
many-valued models $I \in Mod_{\Gamma}$: $~~Biv_{\Gamma} =
\{\Gamma_{I}~|~I \in Mod_{\Gamma} \}$. From the fact that $\Gamma$
is satisfied by every $I \in Mod_{\Gamma}$ we have that for every
$\Gamma_I \in Biv_{\Gamma}$, $\Gamma \subseteq \Gamma_I$, so that
$C(\Gamma) = \bigcap Biv_{\Gamma}$ (the intersection of closed sets
is also a closed set).
 Thus, for $s = (\phi \vdash \psi)$,
$~~~\Gamma \models_m s~~$\\ iff $~~~~\forall I \in
 Mod_{\Gamma} (~ \forall (\phi_i \vdash \psi_i) \in \Gamma(\overline{I}(\phi_i) \sqsubseteq
 \overline{I}(\psi_i))~$ implies $~\overline{I}(\phi) \sqsubseteq \overline{I}(\psi))$\\
 iff $~~~~\forall I \in
 Mod_{\Gamma} (~ \forall (\phi_i \vdash \psi_i) \in \Gamma(\beta(\phi_i \vdash
 \psi_i) = 1)~$ implies $~\beta(\phi\vdash \psi) = 1)$\\
 iff $~~~~\forall v \in
 Mod_{\Gamma} (~ \forall (\phi_i \vdash \psi_i) \in \Gamma((\phi_i \vdash
 \psi_i) \in \Gamma_I)~$ implies $~s \in \Gamma_I)$\\
 iff $~~~~\forall \Gamma_I \in
 Biv_{\Gamma} (~ \Gamma \subseteq \Gamma_I~$ implies $~s \in
 \Gamma_I)$  \\
 iff $~~~~\forall \Gamma_I \in
 Biv_{\Gamma} (~s \in
 \Gamma_I)$ $~~~~$,  because $\Gamma \subseteq \Gamma_I$ for each  $\Gamma_I \in Biv_{\Gamma}$\\
 iff $~~~~~s \in
 \bigcap Biv_{\Gamma} = C(\Gamma)$, that is, $~~~~~$ iff $~~~~~\Gamma \Vdash s$.
 \\$\square$\\
Consequently, in order to define the model-theoretic semantics for a
many-valued logics, we do not need to define the "problematic"
matrices: we are able to use only the many-valued valuations, and
\emph{many-valued models} (i.e., valuations which satisfy all
sequents in $\Gamma$ of
a given many-valued logic $\L$).\\
Differently from the classical logic where a formula is a theorem if
it is true in all models of the logic, here, in  a many-valued logic
$\L$, but specified by a set of sequents in $\Gamma$, for a formula
$\phi \in F(\L)$ that has the same value $x \in X$ (for \emph{any}
algebraic truth-value $x$) for all many-valued models $I \in
Mod_{\Gamma}$, we have that its sequent-based version $\phi \vdash
x$ and $x \vdash \phi$ are theorems; that is, $~~\forall I\in
Mod_{\Gamma} (\overline{I}(\phi) = x)~~$ iff $~~(\Gamma \Vdash (\phi
\vdash x)$ and $\Gamma \Vdash (x \vdash \phi))$. (For instance, in
the case of classical logic, a formula $\phi$ is a theorem iff
$~~(\Gamma \Vdash (\phi \vdash 1)$ and $\Gamma \Vdash (1 \vdash
\phi))$, while $\neg \phi$ is a theorem iff $~~(\Gamma \Vdash (\phi
\vdash 0)$ and $\Gamma \Vdash (0 \vdash \phi))$). But such a value
$x \in A$ does not need to be a designated element $x \in D$, as in
the matrix semantics for a many-valued logic, and it explains why we
do not need the rigid semantic specification by matrix designated
elements.\\ Thus, by  translating  a many-valued logic $\L$ into its
"meta" sequent-based 2-valued logic, we obtain an unambiguous theory
of truth-invariance inference without using the matrices.
\\
\textbf{Remark:} There is also another way to reduce the many-valued
logics into "meta" 2-valued logics, based on the ontological
encapsulation \cite{Majk04on}, where each many-valued proposition
(or many-valued ground atom $p(a_1, ..,a_n)$) is ontologically
encapsulated into 2-valued atom $p_F(a_1, ..,a_n, x)$ (by enlarging
original atoms with new logic variable  whose domain of values is
the set $ A$). Roughly, "$p(a_1, ..,a_n)$ has a value $x$" iff
$p_F(a_1, ..,a_n, x)$ is true).
 In fact such an atom is equivalent to the following formula of
 sequents:
 $(p(a_1, ..,a_n) \vdash x) \wedge (x \vdash p(a_1,
 ..,a_n))$.
 \\$\square$\\
\textbf{Autoreferential possible world semantics:}\\
 Based on
this Gentzen-like sequent deductive system $\G$, or more general
sequent system in \cite{Majk09BS}, with truth-invariance semantics
for the entailment used in the rest of this paper (in Definition
\ref{def:newalg}), we are able to
 define the equivalence relation $\approx_L$ between the formulae of any
 propositional logic based on a  complete distributive lattice  $A$ in order to
 define the Lindenbaum algebra for this logic, $(\L/_{\approx_L}, \sqsubseteq) $,
 where for any two formulae $~\phi, \psi \in \L$,\\ (a)$~~\phi \approx_L
 \psi~~$ iff $~~\phi \vdash \psi$ and $\psi \vdash \phi$, i.e., iff
  $\forall I \in
 \mathbb{V}_m.(\overline{I}(\phi) = \overline{I}(\psi))$. \\
 Thus, each element of the quotient algebra $\L/_{\approx_L}$ is an
 equivalence classes, denoted by $[\phi]$; the partial ordering
 $\sqsubseteq$ is defined by\\ (b)$~~[\phi]\sqsubseteq[\psi]~~$ iff $~~\phi
 \vdash \psi$ (i.e., if $~~\phi \sqsubseteq \psi$).\\
  In particular we will consider an  equivalence class (set
 of all equivalent formulae w.r.t. $\approx_L$) $[\phi]$ that has exactly one
 constant $x \in A$, which is an element of this equivalence class (we abuse a denotation here by denoting by $x$ a formula (logic language constant),
  such that has a constant logic value $x \in A$ for
 every interpretation $I$,  as well), and we
 can use it as the representation element for this equivalence
 class $[x]$. Thus, every formula in this equivalence class has the same
 truth-value as this constant.
 Consequently, we have the injection $i_A:A \rightarrow \L/_{\approx_L}$ between elements in  $(A, \sqsubseteq)$
 and elements in the Lindenbaum algebra, such that for
 any logic value  $x \in A$,  we obtain the equivalence class $[x] = i_A(x)  \in \L/_{\approx_L}$.
 It is easy to extend this injection into an monomorphism between the
 original algebra and this Lindenbaum algebra, by definition of
 correspondent connectives in this Lindenbaum algebra. For example:
 $[x \wedge y] = i_A(x \wedge y) = i_A(x) \wedge_L i_A(y) = [x]
 \wedge_L [y]$, $[\neg x] = i_A(\neg x) = \neg_L i_A(x) = \neg_L[x]$,
 etc..
In an autoreferential semantics we will assume that each equivalence
class of formulae $[\phi]$ in this Lindenbaum algebra corresponds to
one "state - description". In particular, we are interested to the
subset of "state - descriptions" that are \emph{invariant} w.r.t.
many-valued interpretations $I$, so that can be used as the possible
worlds in the Kripke-style semantics for the original many-valued
modal logic. But from the injection $i_A$ we can take for such an
invariant
"state -description" $[x]\in \L/_{\approx_L}$ only its inverse image  $x = i_A^{-1}([x]) \in A$.\\
Consequently, the set of possible worlds in this autoreferential
semantics corresponds to a particular subset of truth values in the
complete lattice $(A, \sqsubseteq)$:
 in this paper we will use the set of join irreducible elements (Birkhoff's representation),
 as semantics based on prime filters, and one more possible world for
the bottom algebraic truth value. Thus, it is from the economical
point of view analogous to the semantics based on prime filters.
\section{A new representation theorem}
%
Based on the considerations in the previous paragraph, we intend to
define an algebraic/relational duality in the way that we do not
need to  define a subset of designated elements $D$ of a many-valued
algebra. Let $I:Var \rightarrow A$ be a given many-valued model of
the logic $\L$, then we can use the \emph{algebraic} model
$(\textbf{A},I)$, instead of o-matrices $(\textbf{A},D)$. Let
$\Gamma$ be a sequent theory for this logic $\L$. The intuitive idea
is to use the \emph{models} $Mod_{\Gamma}$ of the logic $\L$ (notice
that $I \in Mod_{\Gamma}$ is not any valuation for the propositional
variables but is a model, and that the representation theorem is
interesting only for logics that have at least one model, i.e.,
when $Mod_{\Gamma}$ is not empty).\\
In what follows we will consider a poset $A$  of truth values (with
partial ordering $\sqsubseteq$ such that at least for each $x \in A$
we have that $x \sqsubseteq x$) of truth values (nullary operators
of the algebra) for this many-valued logic, and $\{o_i\}_{o_i \in
\O}$ the set of functions $o_i:A^n \rightarrow A$ (with arity $n\geq
1$) assigned to operation names in $\O$ of the logic $\L = (Var, \O,
\Vvdash)$.
 We assume  that the
carrier set of \emph{every} algebra for a logic $\L$ contains also a
set of propositional variables in $Var$, so that the terms of an
algebra $\textbf{A}$ are the terms with variables in $Var$.\\
Consequently, any pair $(\textbf{A},I)$ can be seen as a ground term
algebra
obtained by assigning to $Var$ the values in a model $I$ of $\L$.\\
Thus, the satisfaction relation $\models^a$ will be relative to a
model $I$ of the logic $\L$ instead of the prefixed set of elements
in $D$. For example, in the case of a  logic program $\L$ we can use
the Fitting's 3-valued fixed point operator to obtain its
well-founded 3-valued model. Here we will apply the
\emph{truth-invariance} entailment principle, the idea originally
used to define the inference closure in the bilattice based logics
\cite{MajkC04}, and used recently to develop a new  sequent system
for many-valued logics presented in \cite{Majk09BS} as well: in
these two papers has been described a kind of transformation of the
original many-valued logic into the 'meta' 2-valued logic. The set
of models $Mod_{\Gamma}$ of a given set $\Gamma$ of formulae has to
satisfy
this truth-invariance principle \cite{Majk09BS}: \\
(MV) $~~(\forall \phi \in \Gamma)(\exists x \in A)(\forall I \in
Mod_{\Gamma})(\overline{I}(\phi) = x)$,\\ that is, the value of each
formula in
$\Gamma$ is \emph{invariant} in $Mod_{\Gamma}$. \\
In any case, in the representation theorem framework we are
interested in establishing what is a canonical isomorphic algebra
for a logic $\L$, and its relationship with Kripke relational
structures. So, we can use models $I \in Mod_{\Gamma}$ of a logic
$\L$ only as mean to obtain these results.
 The \emph{algebraic} satisfaction relation
$\models^a$ is defined as follows:
\begin{definition} \label{def:newalg} Let $\L = (Var, \O,
\Vvdash)$ be a logic, $(\textbf{A},I)$ be an algebraic logic model
of a logic $\L$, defined by a mapping $I:Var \rightarrow A$, and
$\phi \in F(\L)$, and $\overline{I}:F(\L) \rightarrow A$ be its
unique standard extension to all formulae in a language $\L$. Let
$\M$ be a class of algebraic models.\\ We define the
relation $\models^a$ as follows $(x \in A)$:\\
1. $(\textbf{A},I);x\models^a\phi~$ iff $~x = \overline{I}(\phi)$,\\
2. $\M;x \models^a\phi~$ iff $~(\textbf{A},I);x\models^a~\phi~$
for every $~(\textbf{A},I) \in \M$.\\
We define the entailment relation of a logic $\L$ by:  for every
$~\phi \in F(\L), x \in A$,
 $~\L;x \Vvdash \phi$ iff $~\M;x \models^a\phi$.
\end{definition}
Notice that in this definition, analogous to Definition
\ref{def:matr}, we do not use the set of designated values $D$, and
we are able to determine which set of formulae is deduced \emph{for
each} algebraic logic value $x \in A$. It is a generalization of
classical deduction, where $\L \Vvdash \phi$ is equivalent to this
new definition $\L;1 \Vvdash \phi$, and $\L \Vvdash \neg\phi$ is
equivalent to $\L;0 \Vvdash \phi$ (i.e., $\L;1 \Vvdash \neg\phi$).
The inference  of $\phi$ defined by Definition \ref{def:matr}, based
on set $D$ of designated values, can be expressed from this more
accurate definition above by  $\bigvee_{x \in D} \L;x \Vvdash \phi$.
Thus, this new entailment relation $\Vvdash$ given by Definition
\ref{def:newalg} is more powerful and more general than the
entailment relation of $\L$ given by Definition \ref{def:matr}.
\\Notice that if $\Gamma$ is a sequent theory for $\L$, then $~\L;x \Vvdash \phi$ iff
$~\forall I \in Mod_{\Gamma}(\overline{I}(\phi) = x)$,
 that is, in the case of the sequent system presented in Section 2, $\Gamma \Vdash (\phi
\vdash x)$ and  $\Gamma \Vdash (x \vdash \phi)$. Consequently,
$\Vvdash$ satisfies the truth-invariance principle (MV).
 Now we can introduce a \emph{new definition} of the
algebraic/relational duality, as follows:
\begin{definition} \label{def:abs1}
Let $\M$ be a class of all algebraic models for a given logic $\L$.
Assume that there exists a class $\K_{\M}$ of Kripke-models of a
logic $\L$, $(K,I_K)\in \K_{\M}$, with a Kripke-frame $K = (1_K,
\{R_j\}_{j \leq n})$ where $1_K$
is the set of possible worlds, $\{R_j\}_{j \leq n}$ a finite set of
accessibility relations between them (relational structure), with a
mining mapping $I_K:Var \rightarrow \P(1_K)$,  such that there
exists a mapping $\mathbb{D}:\M \rightarrow \K_{\M}$, with $\K_{\M}
= \{\mathbb{D}(M) ~|~ M = (\textbf{A},I) \in \M\}$, and there exists
a mapping $\mathbb{E}:\K_{\M} \rightarrow \M$ such that:
\begin{description}
                            \item[(i)] for every \emph{Kripke} model $M_K = (K,I_K) \in \K_{\M}$ of
                            $\L$,
                            the $~  (\textbf{A}_K, I_K)= \mathbb{E}((K,I_K))  \in
                            \M$ is an algebraic model of $\L$, where $\textbf{A}_K = (\P(1_K), \{o_K\}_{o \in \O})$ is an algebra
                            of subsets of the support $1_K$ of $K$;
                            \item[(ii)] for every \emph{algebraic} model $M = (\textbf{A},I)
                            \in \M$ of $\L$, the $(K,I_K) = \mathbb{D}(M)$ is a Kripke model over a set
                            $1_K$, so that, if $\mathbb{E}(\mathbb{D}(M))
                            =(\textbf{A}_{K}, I_{K})$ then there is an
                            monotone injection mapping $~i_n:A \hookrightarrow A_{K}$, between truth values of algebras $\textbf{A} = (A, \sqsubseteq, \{o\}_{o \in
                            \O})$
                             and $\textbf{A}_{K} = (A_{K},\subseteq, \{o_{K}\}_{o \in \O})$, where $A_K = \P(1_K)$,
                              such that  $~I_{K} = i_n \circ I$ and $D_K =
                              \{i_n(x)~|~x \in A \}$.
\end{description}
A representation is \verb"autoreferential" when $1_K \subseteq A$.
\end{definition}
 \textbf{Example 3}: \\
 Let us consider the two following \emph{autoreferential} representations:\\ Case A: Let us consider the standard propositional logic $\L = (Var, \O,
\Vvdash)$, where $\O = \{\wedge, \sim\}$ and its simple Boolean
algebra $\textbf{A} = (A,\sqsubseteq, \{\wedge, \sim\})$, where $A =
\textbf{2} = \{0,1\}$ with logic operators 'and', $\wedge$, and
logic negation $\sim$ respectively, with $0 \sqsubseteq 1$. Let us
take $1_K = A = \{0,1\}= \textbf{2}$, so that the canonical
extension of the Boolean algebra $\textbf{A}$ is the powerset
algebra $(\P(\{0,1\}), \subseteq, \{\bigcap, \neg\})$, with
inclusion homomorphism $i_n:(A, \sqsubseteq, \{\wedge, \sim\})
\rightarrow (\P(\{0,1\}), \subseteq, \{\bigcap, \neg \})$, which
\emph{preserves ordering}, such that for its bottom and top
elements hold, $0_K = i_n(0) =  \{0\}, 1_K = i_n(1) = \{0,1\}$.\\
The negation algebraic operator $\neg$ is defined by $\neg X = X
\Rightarrow \{0\}$, where the operator (implication) $\Rightarrow$
is defined by $X\Rightarrow Y = \bigcup\{Z\in \P(\{0,1\})~|~Z
\bigcap X \subseteq Y\}$, for any $ X,Y \in
\P(\{0,1\})$.\\
Notice that $\neg$ is not an involution in $\P(\{0,1\})$, because
$\neg \neg(\{1\}) = \{0,1\} \neq  \{1\}$. But it is an involution
negation operator for the  \emph{subalgebra} of this canonical
extension, $(D_K, \{\bigcap, \neg \})$, where $D_K =\{\{0\}
,\{0,1\}\} \subset \P(\{0,1\})$, isomorphic to  algebra $\textbf{A}$
and
defined by the image of the inclusion  $i_n$.\\
Case B:  Let us consider the 4-valued Belnap's distributive
bilattice $A = \{f,\bot, \top, t\}$ with $\bot$ for \emph{unknown}
and $\top$ for \emph{inconsistent} logic value, $f = 0, t = 1$ are
bottom and top values w.r.t the \emph{truth} ordering  $0
\sqsubseteq \bot, ~0 \sqsubseteq \top, ~\bot \sqsubseteq 1, ~\top
\sqsubseteq 1$ and $\bot \bowtie \top$. It is the smallest
many-valued logic capable of dealing with incomplete (unknown) and
inconsistent logics. In this case we can take $1_K = \{0,\bot,
\top\} \subset A$, with monotone injection $i_n:(A, \sqsubseteq,
\{\wedge, \vee\}) \hookrightarrow (\P(1_K), \subseteq,
\{\bigcap,\bigcup\})$ such that: $0_K = i_n(0) = \{0\}, i_n(\bot) =
\{0, \bot\},  i_n(\top) = \{0, \top\}, 1_K = i_n(1) = \{0, \bot,
\top\}$, i.e., $D_K = \{\{0\}, \{0, \bot\}, \{0, \top\}, \{0, \bot,
\top\}\}$.
\\$\square$\\
In this new definition we replaced the old duality Algebras -
Relational structures by the semantic duality Algebraic models -
Kripke models of a logic $\L$.\\
Notice that in the definition above we do not require  the injection
$i_n$ to be an injective \emph{homomorphism}, as in the assumption
\ref{def:abs}, but we require  that the following diagram commutes
(here $id_{A_K}$ is the identity mapping for $A_K$):
 \begin{diagram}
Var  & \rTo^{I} & A &\lTo^{\overline{I}} & F(\L)  \\
 \dTo_{I_{K}} & & \dTo_{i_n} & &\dTo_{\overline{I}_{K}}      \\
  A_K & \rTo^{id_{A_K}} &   A_{K}   &     \lTo^{id_{A_K}}    &A_K
\end{diagram}
\begin{definition}  \label{def:newrelatsem} Assume that $\M$ and $\K_{\M}$ satisfy the
assumptions  in \ref{def:abs1}. Let $(K,m) \in \K_{\M}$, $~1_K$ be
the support of $K$,with
 $m:Var \rightarrow \P(1_K)$ and
$~\overline{m}:F(\L) \rightarrow \P(1_K)$ be the unique extension of
$~m$ for all formulae in $F(\L)$. Let  $y \in 1_K$ and $\phi \in F(\L)$, then:\\
1. $~~~(K,m) \models_{y}~\phi~~~$ iff  $~~~y \in \overline{m}(\phi)$;\\
2. $~~~(K,m) \models~\phi~~~$ iff $~~~ \overline{m}(\phi) \in D_K$;\\
3. $~~~\K_{\M} \models~\phi~~~$ iff $~~~~\forall(K,m_i),(K,m_j) \in
\K_{\M} (\overline{m_i}(\phi) = \overline{m_j}(\phi) \in D_K)$.
\end{definition}
The following theorem is the basic result for the next
representation theorem, and shows that from Definition
\ref{def:abs1} the new relational inference $\models$ is sound and
complete w.r.t. the algebraic inference $\models^a$.
\begin{theo} Assume that $\M$ and $\K_{\M}$ satisfy the assumptions in  \ref{def:abs1}.
Then, for every $\phi \in F(\L)$,\\ if $~\M;x \models^a~\phi~~$ then
$~~\K_{\M} \models~\phi$ with $i_n(x) = \overline{m}(\phi)$ for any
$(K,m) \in \K_{\M}$. The converse also holds.
\end{theo}
\textbf{Proof}: Assume that  $\M$ and $\R$ satisfy the assumptions
in \ref{def:abs1} and $\phi \in F(\L)$ such that $~(\textbf{A},I);x
\models^a~\phi$, i.e., $x = \overline{I}(\phi)$. Let $(K,I_K) \in
\K_{\M}$, $\overline{m}:F(\L) \rightarrow A_K$ be the unique
extension of $~m:Var \rightarrow \P(1_K)$. By (C)(i) we have that
$~\mathbb{E}((K,I_K)) = (\textbf{A}_K,I_K) \in \M$, with $A_K =
\P(1_K)$.
 Then, from $\overline{I}_K = i_n\circ \overline{I}$
it holds that $\overline{I}_K (\phi)= i_n(\overline{I}(\phi)) =
i_n(x)$ and $~(\textbf{A}_K,I_K);i_n(x) \models^a \phi$. That is,
for any $~g:Var \rightarrow A_K$, thus also for $m$, $~
\overline{m}(\phi) = \overline{I}_K(\phi) = i_n(x) \in D_K$,
 in the way that  $(K,m) \models \phi$. It is valid for any  $(K,I_K) \in \K_{\M}$, thus $~~\K_{\M} \models \phi$.\\
 Let $\phi \in F(\L)$, with $~\K_{\M}\models \phi$. Then for any $~M =
(\textbf{A},I) \in \M$, we have $~(K,I_K) = \mathbb{D}(M) \in
\K_{\M}$.
 Since for any $~(K,m) \in \K_{\M}~$ we know that
$~(K,m) \models \phi$, that is, for any $m:Var \rightarrow A_K~$
(thus for $I_K$ also), $\overline{m}(\phi) = i_n(x) \in D_K$ for
some $x \in A$, so that $~\overline{I_K}(\phi) = i_n(x) \in D_K$,
 and from the fact that $I_K = i_n\circ I$, we can
take $x = I(\phi)$. Thus $(\textbf{A},I);x\models^a \phi$.\\
 Since it holds for any
$~M =(\textbf{A},I) \in \M$, we obtain  $~~\M;x \models^a~\phi$.\\
$\square$
\begin{coro} Let $\L = (Var, \O,
\Vvdash)$ be sound and complete  logic w.r.t. a class $\M$ of
algebraic models. Assume that there exists a class $\K_{\M}$ such
that the assumption in \ref{def:abs1} holds. Then $\L$ is sound and
complete w.r.t. the class $~\K_{\M}$, which can be regarded as a
class of Kripke-style models.
\end{coro}
From this corollary we are able to define a direct duality between
algebraic and Kripke-style semantics for a logic $\L$
\begin{diagram}
 & \M   \rTo^{\mathbb{D}}& \K_{\M} & \rTo^{\mathbb{E}} \M &
\end{diagram}
\begin{theo}\label{def:RepresTh}\textsc{Representation Theorem:} Assume that $\M$ and $\R$ satisfy the assumptions in \ref{def:abs1}.
Injective mapping $i_n$ can be extended to the injective
homomorphism $~i_n:(\textbf{A},I)\hookrightarrow
(\textbf{A}_K,I_K)~$, where $(K,I_K) = \mathbb{D}(\textbf{A},I)$.
Thus, the \verb"dual" representation of the algebra $\textbf{A}$ is
the subalgebra of $\textbf{A}_K$ defined by image of the
homomorphism $~i_n$.
\end{theo}
\textbf{Proof}: It comes from the fact that $I$ and $I_K$ are the
homomorphisms between O-algebras. So we can show it by structural
induction on the formulae in $F(\L)$. For example, for a formula
composed by conjunction, $\phi \wedge \psi$, with $x =
\overline{I}(\phi), y = \overline{I}(\psi)$, we have that, $i_n(x
\wedge_A y) =\\ = i_n(\overline{I}(\phi) \wedge_A
\overline{I}(\psi)) = i_n(\overline{I}(\phi \wedge \psi)) $, $~~~~$
from the homomorphic property of
$\overline{I}$\\
$ = (i_n \circ \overline{I})(\phi \wedge \psi)) =
\overline{I}_K(\phi \wedge \psi)  $, $~~~~~~$
from the commutativity of (C)(ii)\\
$=\overline{I}_K(\phi) \wedge_K \overline{I}_K(\psi) $ $ = i_n(
\overline{I}(\phi)) \wedge_K i_n(
 \overline{I}(\psi))\\ = i_n(x) \wedge_K i_n(y)$.
 Thus, we obtained that the homomorphism holds for the restriction of
 $~i_n$ to the image of $I$, but it is generally valid for \emph{any}
 $I$.\\$\square$\\
 \textbf{Example 4}: (The continuation of Example 3)\\ Let us consider now the algebraic models for $\L$, based on the
 Boolean algebra, $(\textbf{A}, I) \in \M$, where $I:Var
\rightarrow \textbf{2}$ is the interpretation for propositional
variables in $Var$, and on  its canonical extension
$(\textbf{A}_K,I_K) = \mathbb{E}(\mathbb{D}((\textbf{A}, I)))$,
 where $(K,I_K) = \mathbb{D}((\textbf{A}, I)), ~~\textbf{A}_K = (\P(\{0,1\}),
\subseteq, \{\bigcap, \neg\})$ .\\
We have that for any $p \in Var$, $I(p) = 1~$ iff $~I_K(p) =
\{0,1\}$ and $~I(p) = 0~~$ iff $~~I_K(p) = \{0\}$.\\
We do not have  any modal operator in these algebras, thus the frame
$K \in (K,I_K) = \mathbb{D}((\textbf{A}, I))$ has the set of only
two possible worlds equal to $1_K = \textbf{2} = \{0,1\}$ and an
empty accessibility relation, that is $K = (\{0,1\}, \{\})$.
\section{Autoreferential representation for complete distributive lattices}
In  Examples 3 and 4 we have shown  the cases for  this new
definition of representation theorem, based on models of a logic
$\L$, which define only relational structures $K \in (K,I_K) =
\mathbb{D}((\textbf{A},I))$, with a set of possible worlds (support)
equal to
the set $1_K \subseteq A$. \\
In the rest of this paper we will consider the subclass of complete
latices in which each lattice of truth values $(A, \sqsubseteq,
\wedge, \vee)$  is \emph{isomorphic} to the complete sublattice of
the powerset lattice $(\P(A),\subseteq, \bigcap, \bigcup)$.
Consequently, we will consider the cases when there exists the
subset $S = C_L(\P(A)) \subseteq \P(A)$, closed under intersection
$\bigcap$ and union $\bigcup$, with the isomorphism $i_s:(A,
\sqsubseteq, \wedge, \vee) \simeq (C_L(\P(A)),\subseteq, \bigcap,
\bigcup)$, so that we obtain the inclusion map $i_n = ~\subseteq
\circ~ i_s:A \hookrightarrow A_K = \P(1_K) $ as required in Definition \ref{def:abs1}.\\
For such a subclass of complete lattices we will obtain that the
carrier set $A$, of the many-valued logic algebra $\textbf{A}$, is
the set of possible worlds for the Kripke frame for the  dual
relational representation of the algebraic semantics: this is an
\emph{autoreferential assumption} \cite{Majk06ml}. The relational
semantic of other modal operators
of the algebra $\textbf{A}$ will be obtained successively
by  a correct definition of the accessibility relations of the
Kripke frame.\\
It is well known that any complete lattice $A$ has the following
property: each (also infinite) subset $X$ of $A$ has the least upper
bound (supremum) denoted by $\bigvee X$ (when $X$ has only two
elements, the supremum corresponds to the join operator $\vee$), and
the greatest lower bound (infimum) denoted by $\bigwedge X$ (when
$X$ has only two elements  the infimum corresponds to the meet
operator $\wedge$). Thus, it has the bottom element $0 = \bigwedge A
\in A$, and the top element $1 = \bigvee A \in A$. The cardinality
of the set of hereditary subsets of $A$ is generally greater than
the cardinality of the lattice $A$. But in what follows we will
consider the class of complete \emph{distributive} lattices $A$, for
which we are able to define an isomorphism \cite{Birkh40} between
the original lattice $A$ and the particular collection $A^+$ of
hereditary subsets of $A$. Thus, in each  distributive lattice we
are able to define the implication and negation logical operators
based on relative pseudocomplement and pseudocomplement relatively,
i.e., $a\rightharpoonup b = \bigvee S$,  $S =\{x\in A | x \wedge a
\sqsubseteq b\}$ and  $\sim a = a
\rightharpoonup 0$. \\
 \textbf{Example 5:} Many-valued logics for approximate truth enriched by approximation of \emph{unknown} and \emph{inconsistent} information:
   The class of poset
lattices can also be used  for enabling standard fuzzy logic over
the closed interval $[0,1]$ of reals, with whenever $x \leq y$ then
$x \sqsubseteq y$ where $\leq$ is the standard ordering of numbers,
used for approximation of the truth value, with ability to consider
incomplete (unknown) and mutually inconsistent information as well.
For example, let us consider an enriched fuzzy logic with the set of
truth values in $[0, .5-\triangle] \bigcup[ .5+\triangle,1]\bigcup
\{\bot, \top\}$ where $.5-\triangle \sqsubseteq \bot \sqsubseteq
5+\triangle$ and $.5-\triangle \sqsubseteq \top \sqsubseteq
5+\triangle$, for an sufficiently small value $\triangle < .5$. In
the simplest case we can substitute $.5$ value with two unrelated
values $0.5^- = \bot$ and $0.5^+ = \top$. This more expressive fuzzy
logic we will denominate PO-fuzzy logic. This enrichment of the
fuzzy logic is obtained by replacement of the closed subinterval
$[x-\triangle_x, x+\triangle_x]$ of reals by the discrete Belnap's
bilattice $\{x-\triangle_x, \bot_x, \top_x, x+\triangle_x\}$ for an
enough small $\triangle_x$. We are able also to repeat such an
operation for a number of such replacements for different values for
$x \in [0,1]$, with the family of unknown and inconsistent values
such that if $x \leq y$ then $\bot_x \sqsubseteq \bot_y$ and $\top_x
\sqsubseteq \top_y$, and $x+\triangle_x < y-\triangle_y$, in order
to have not only the fuzzy approximation of truth values, but also
the approximations of unknown and
inconsistent values. Each such an enrichment is a distributive lattice.\\
Obviously, each finite or infinite many-valued logic with total
ordering can be enriched by the family of values $\bot_x$ and
$\top_x$,  for the approximations of the unknown and inconsistent
values, in order to be able to deal with any kind of incomplete and
inconsistent information.
\\$\Box$\\
From the Birkhoff's representation theorem \cite{Birkh40} for
distributive lattices, every finite (thus complete) distributive
lattice is isomorphic to the lattice of lower sets of the poset of
join-irreducible elements. An element $x \neq 0$ in $A$ is a
join-irreducible element iff $x = a \vee b$ implies $x = a$ or $x =
b$ for any $a,b \in A$. Lower set (down closed) is any subset $Y$ of
a given poset $(A,\sqsubseteq)$ such that, for all elements $x$ and
$y$, if $x \sqsubseteq y$ and $y \in Y$ then $x\in Y$.
\begin{propo} \cite{Birkh40} \textsc{0-Lifted Birkhoff isomorphism:}  \label{def:canisomor}
Let $A$ be a complete distributive lattice, then we define the
following mapping $~~\downarrow^+:A \rightarrow \P(A)$: for any $x
\in A$, $~~\downarrow^+ x ~ = ~ \downarrow x ~\bigcap~\widehat{A}$,
where \\$\widehat{A} = \{y~|~y \in A~$ and $~y~$ is
join-irreducible $\} \bigcup \{0 \}$. \\
 We define the set $A^+ = \{\downarrow^+ a ~| a \in A~\} \subseteq
\P(A)$,  so that $\downarrow^+ \bigvee = id_{A^+}:A^+ \rightarrow
A^+$  and $\bigvee \downarrow^+ = id_A:A \rightarrow A$.  Thus, the
operator $\downarrow^+$ is inverse of the supremum operation
$\bigvee:A^+ \rightarrow A$. The set $(A^+, \subseteq)$ is a
complete lattice, such that there is the following 0-lifted Birkhoff
isomorphism $~~\downarrow^+:(A, \sqsubseteq, \wedge, \vee) \simeq
(A^+,\subseteq, \bigcap, \bigcup)$.
\end{propo}
\textbf{Proof}: Let us show the homomorphic property of
$\downarrow^+$:\\
$\downarrow^+ (x\wedge y) = \downarrow(x\wedge
y)~\bigcap~\widehat{A} = (\downarrow x \bigcap \downarrow
y)~\bigcap~\widehat{A} =\\= (\downarrow x ~\bigcap~\widehat{A})
\bigcap (\downarrow y)~\bigcap~\widehat{A}) =  \downarrow^+ x
~\bigcap~\downarrow^+ y$, $~~$ and \\
$\downarrow^+ (x\vee y) = \downarrow(x\vee y)~\bigcap~\widehat{A} =
(\downarrow x \bigcup \downarrow y)~\bigcap~\widehat{A} = \\ =
(\downarrow x ~\bigcap~\widehat{A}) \bigcup (\downarrow
y)~\bigcap~\widehat{A}) =  \downarrow^+ x ~\bigcup~\downarrow^+
y$. \\
The isomorphic property holds from Birkhoff's representation
theorem.
\\ $\square$ \\
The name lifted here is used to denote the difference from the
original Birkhoff's isomorphism. That is, we have that for any $x
\in A$, $0 \in \downarrow^+ x ~$, so that $\downarrow^+ x$ is never
empty set
(it is lifted by bottom element $0$).\\
Notice that $(A^+,\subseteq, \bigcap, \bigcup)$ is a
\emph{subalgebra} of the powerset algebra $(\P(A), \subseteq,
\bigcap, \bigcup)$.\\\\
 \textbf{Example 6}:  Belnap's bilattice in the Example 5, is a
 distributive lattice w.r.t. the $\leq_t$ ordering, with two join-irreducible elements
  $\bot$ and $\top$, so that $\widehat{\B} = \{0, \bot, \top\}$.
 In this case we have that $\downarrow^+ 1 = \downarrow^+ (\bot \vee \top) =
 \downarrow^+
\bot \bigcup \downarrow^+ \top= \downarrow \bot \bigcup \downarrow
\top = \{0, \bot, \top\} = \widehat{\B} ~~\neq~~ \downarrow 1 = \B$.
\\$\square$\\
 It is easy to verify that $\downarrow^+ 0 = \{0\}$ is the bottom element
in $ A^+$.\\
\textbf{Remark}: For a many-valued logic with distributive complete
lattice of truth values we have that $A_K = \P(1_K) \subseteq
\P(A)$, with $1_K = \widehat{A}$ and $D_K = A^+$, and the injective
homomorphism $\downarrow^+:A \rightarrow \P(A)$ corresponds to the
injective homomorphism $~i_n:(\textbf{A},I)\hookrightarrow
(\textbf{A}_K,I_K)~$ in the representation theorem
\ref{def:RepresTh}. Thus, the \verb"dual" representation of this
algebra (in this case a distributive complete lattice) $\textbf{A}$
is the subalgebra
 $(A^+,\subseteq, \bigcap,\bigcup)$ of $\textbf{A}_K$, defined by the image of the homomorphism $~~i_n = \downarrow^+$.\\
 Based on these results we are able to extend the complete distributive lattices with other
 unary algebraic operators $\{o_i\}_{i \in N}:A \rightarrow A$ and binary operators $\{\otimes_i\}_{i \in N}:A\times A \rightarrow A$ in order to obtain a class of algebras
 $((A,\sqsubseteq, \wedge, \vee), \{o_i\}_{i \in N}, \{\otimes_i\}_{i \in N})$,
 with the following set-based canonical representation:
\begin{propo} \textsc{Canonical representation:} \label{prop:canonic}
 Let $ \textbf{A} =((A,\sqsubseteq, \wedge, \vee),  \{o_i\}_{i \in N}, \{\otimes_i\}_{i \in
N})$ be a complete distributive lattice-based algebra. \\ We define
its
 canonical representation by the algebra $ \textbf{A}^+ =
((A^+,\subseteq, \bigcap, \bigcup), \{o^+_i\}_{i \in N},
\{\otimes^+_i\}_{i \in N})$, such that,\\ $o^+_i = (\downarrow^+ o_i
\bigvee):A^+ \rightarrow A^+$ and $\otimes^+_i = (\downarrow^+
\otimes_i \bigvee):A^+ \times A^+ \rightarrow A^+$ are the unary and
binary  operators over sets in $A^+$.
\end{propo}
\textbf{Proof:} We have that for any $x,y \in A$, $\downarrow^+
o_i(x) = \downarrow^+ o_i ( \bigvee \downarrow^+)(x) = (\downarrow^+
o_i  \bigvee) \downarrow^+ x = o^+_i (\downarrow^+ x)$ and
$\downarrow^+ \otimes_i(x,y) = \otimes^+_i (\downarrow^+ x,
\downarrow^+ y)$. Thus, $\downarrow^+$ is an isomorphism
$\downarrow^+: \textbf{A} \simeq \textbf{A}^+$. \\ $\square$ \\
\textbf{Example 7}: Let us consider the binary implication operator
$\otimes_i$ equal to the relative pseudocomplement $\rightharpoonup$
over a complete distributive lattice. Then, we have that
$(\downarrow^+ x) \rightharpoonup^+ (\downarrow^+ y) = \otimes^+_i
(\downarrow^+ x, \downarrow^+ y) = \downarrow^+ \otimes_i(x,y) =
\downarrow^+(x\rightharpoonup y) = \downarrow^+ (\bigvee\{z~|~z
\wedge x \sqsubseteq y\} = \bigcup \{\downarrow^+ z~|~z \wedge x
\sqsubseteq y\} = $ (from the homomorphism $\downarrow^+ $ w.r.t.
the join operator of this lattice)\\ $ = \bigcup \{\downarrow^+
z~|~\downarrow^+(z \wedge x) \subseteq \downarrow^+ y\} = $ (from
$\downarrow^+v \subseteq \downarrow^+w$ iff $v \sqsubseteq w$)\\ $ =
\bigcup \{\downarrow^+ z~|~\downarrow^+ z \bigcap \downarrow^+ x)
\subseteq \downarrow^+ y\} = $ (from the homomorphism $\downarrow^+
$ w.r.t. the meet operator of this lattice)\\ $ = \bigcup \{S \in
A^+~|~S \bigcap \downarrow^+ x) \subseteq \downarrow^+ y\} $. \\That
is, we obtain that the correspondent operator $\otimes^+_i:A^+
\times A^+ \rightarrow  A^+$ is a relative pseudocomplement for the
lattice $A^+$.
\\ $\square$ \\
It is well known that the standard unary \emph{existential modal}
operators are homomorphisms between join semilattices,
$o_i:(A,\sqsubseteq, \vee) \rightarrow (A,\sqsubseteq, \vee)$, and
modal negation operators $\widetilde{o_i}:(A,\sqsubseteq, \vee)
\rightarrow (A,\sqsubseteq, \vee)^{OP}$, where the dual join
semilattice $(A,\sqsubseteq, \vee)^{OP}$ has $\sqsubseteq^{OP} =
\sqsupseteq$, and $\vee^{OP} = \wedge$. For the \emph{normal} modal
logics they are monotone ($x\sqsubseteq y$ implies
$o_i(x)\sqsubseteq o_i(y)$, and $\widetilde{o_i}(x)\sqsubseteq^{OP}
\widetilde{o_i}(y)$)) , additive ($o_i(x \vee y) = o_i(x) \vee
o_i(y)$, and ($\widetilde{o_i}(x \vee y) = \widetilde{o_i}(x)
\vee^{OP} \widetilde{o_i}(y) = \widetilde{o_i}(x) \wedge
\widetilde{o_i}(y)$),
 and normal ($o_i(0) = 0$, and $\widetilde{o_i}(0) = 0^{OP} = 1$).
 Now we are able to
show that for any algebraic model $M = (\textbf{A}, I)$, of a logic
$\L$ with relative pseudocomplement for implication and a number of
unary modal operators, there exists the correspondent Kripke model $
\M_K = \mathbb{D}(M) = (K,I_K)$. In what follows we denote by
$\Rightarrow$ the \emph{logic} connective for implication,
correspondent to the \emph{algebraic} relative pseudocomplement
$\rightharpoonup$, by $\diamondsuit_i$ the existential modal
connective for the algebraic additive operator $o_i$, and by
$\neg_i$ the logic negation modal connective for the algebraic
additive negation operator $\widetilde{o_i}$, so that for the
homomorphism (valuation) $\overline{I}:F(\L)\rightarrow A$ holds
that $\overline{I}(\phi \Rightarrow \psi) = \overline{I}(\phi)
\rightharpoonup \overline{I}(\psi)$, $\overline{I}(\diamondsuit_i
\phi) = o_i(\overline{I}(\phi))$ and $\overline{I}(\neg_i \phi) =
\widetilde{o_i}(\overline{I}(\phi))$.\\
Notice that if we denote by $\downarrow_A:\widehat{A} \rightarrow
\P(\widehat{A}) $ the restriction of $\downarrow: A
\rightarrow\P(A)$ to the subset of join-irreducible elements
$\widehat{A} \subseteq A$, then for any $x \in \widehat{A}$ we
obtain\\
(*)  $~~\downarrow^+ x = \downarrow_A x = \{y \in
\widehat{A}~|~y \sqsubseteq x\} ~~\in A^+ \subseteq \P(\widehat{A})$. \\
Consequently, in the next Kripke-style definition for modal
many-valued logics we will use the set $\widehat{A}$, of
join-irreducible elements in
$A$, for the set of possible worlds.\\
Now we will define the standard accessibility relation for any given
additive normal modal operator $o_i$ and negation modal operator
$\widetilde{o_i}$.
\begin{definition} \label{def:frame} Let $o_i:(A,\sqsubseteq, \vee) \rightarrow
(A,\sqsubseteq, \vee)$ and  negation operator
$\widetilde{o_i}:(A,\sqsubseteq, \vee) \rightarrow (A,\sqsubseteq,
\vee)^{OP}$, be  the additive normal modal  operators. Then we
define the accessibility
relation for $o_i$ by\\
$\R_i =  ~\{(x,y)~|~y  \in \widehat{A}~$, and $~x~\in \downarrow^+
o_i(y)\}$,\\
and the incompatibility relation for $\widetilde{o_i}$ by\\
$\widetilde{\R_i} =  ~\{(x,y)~|~ z \in A, ~y  \in \downarrow^+ z~$,
and $~x~\in \downarrow^+ \widetilde{o_i}(z)\}$.
\end{definition}
\textbf{Remark:} More about a hierarchy of negation operators for
complete lattices  and their relational semantics can be found in
\cite{Majk06ml}. This semantics is based on the Birkhoff concept of
\emph{polarity} \cite{Birkh40}: If $(X, R )$ is a set with a
particular relation on a set $X$, $R \subseteq X \times X$,  with
mappings
$\lambda:\P(X)\rightarrow \P(X)^{OP}, \varrho:\P(X)^{OP}\rightarrow \P(X)$, such that for subsets $U,V \in \P(X)$, \\
$\lambda U = \{x \in X~|~\forall u \in U.( (u,x) \in R) \},
~~\rho V = \{x \in X~|~\forall v \in V. ((x,v) \in R )\}$,\\
where the powerset $\P(X)$ is a  \emph{poset} with bottom element
empty set $\emptyset$ and top element $X$, and $\P(X)^{OP}$ is its
dual (with  $\subseteq^{OP}$ inverse of $\subseteq$). Then we have
an induced Galois connection $\lambda \dashv \rho$, i.e., $\lambda U
\subseteq^{OP} V~$ iff $~U \subseteq \rho V$. The additive modal
operator $\lambda$ is a set-based correspondent operator for the
modal negation operator $\widetilde{o_i}$, when we consider the
relation $R$ as an \emph{incompatibility} (or "perp") relation
$\widetilde{\R_i}$  in
 Definition \ref{def:frame} for this modal negation operator, and
$~~\lambda U = \{x \in X~|~\forall u (u \in U$ implies $ (u,x) \in
\widetilde{\R_i}) \}~$, which will be used for the relational
Kripke-style semantics of
modal negation operators in what follows.\\
 \textbf{Example 8}:
 The smallest \emph{nontrivial} distributive bilattice is Belnap's 4-valued
bilattice  ~\cite{Beln77} $\B = \{t,f,\bot, \top\}$ where $t$ is
\emph{true}, $f $ is \emph{false}, $\top$ is inconsistent (both true
and false) or \emph{possible }, and $\perp$ is \emph{unknown}. As
Belnap observed, these values can be given two natural orders:
\emph{truth} order, $\leq_t$, and \emph{knowledge} order, $\leq_k$,
such that $f \leq_t \top \leq_t t$, $~f \leq_t \bot \leq_t t$, $\bot
\bowtie_t \top$ and $\bot \leq_k f \leq_k \top$, $~\bot \leq_k t
\leq_k \top$, $f \bowtie_k t$. That is, bottom element $0$ for
$\leq_t$ ordering is $f$, and for $\leq_k$ ordering is $\bot$,
 and top element $1$ for $\leq_t$ ordering is $t$, and for $\leq_k$ ordering is $\top$.
Meet and join operators under $\leq_t$ are denoted $\wedge$ and
$\vee$; they are natural generalizations of the usual conjunction
and disjunction notions. Meet and join under $\leq_k$ are denoted
$\otimes$ and $\oplus$, such that hold: $~f \otimes t = \bot$, $f
\oplus t =\top$, $\top\wedge \bot
= f$ and $\top \vee \bot = t$.\\
There is a natural notion of the \emph{bilattice truth negation},
denoted $\neg$, (reverses the $\leq_t$ ordering, while preserving
the $\leq_k$ ordering): switching $f$ and $t$, leaving $\bot$ and
$\top$, and corresponding knowledge negation (\emph{conflation}),
denoted $-$, (reverses the $\leq_k$ ordering, while preserving the
$\leq_t$ ordering), switching $\bot $ and $\top$, leaving $f$ and
$t$. These two kinds of negation commute: $-\neg x = \neg-
x$ for every member $x$ of a bilattice.\\
In what follows we will use the \emph{relative pseudocomplements},
defined by $x \rightharpoonup y = \bigvee\{z~|~z \wedge x \leq_t
y\}$, and \emph{pseudocomplements}, defined by $\neg_t x = ~\sim x =
x \rightharpoonup f$ (and, analogously, for $\leq_k$ ordering, $x
\rightharpoondown y$ and  $\neg_k x = x \rightharpoondown \bot$).
\\
 The conflation is a monotone function that preserves all
finite meets (and joins) w.r.t. the lattice $(\B, \leq_t)$, thus it
is the universal (and existential, because $- = \neg - \neg$) modal
many-valued operator: "it is believed that" for a bilattice (as in
ordinary 2-valued logic, the epistemic negation is composition of
strong negation $\neg_t$ and this belief operator, $\neg = \neg_t
-$), which extends the 2-valued belief of the autoepistemic
logic as follows:\\
1. if $A$ is true than "it is believed that A", i.e., $-A$, is
  true;\\
 2. if $A$ is false than "it is believed that A" is
  false;\\
3. if $A$ is unknown than "it is believed that A" is
  inconsistent: it is really inconsistent to believe in something
  that is unknown;\\
 4.  if $A$ is inconsistent (that is \emph{both} true and false) than "it is believed that A" is
  unknown: really, we can not tell nothing about believing in something that is
  inconsistent.\\
  \textbf{Remark:} Notice that the knowledge negation operator $-$ is \emph{normal} additive modal
  operator w.r.t. the $\leq_t$ ordering. As we will see in the next
  definition, its dual is truth negation $\neg$ which is
  a normal modal operator w.r.t. the $\leq_k$ ordering.
 Thus, in the case of the believe
(conflation) modal operator $o_i = -$ in Belnap's bilattice,
$\widehat{A} = \{f ,\bot,\top \}$,  such that $-f = f, ~-t = t, ~
-\bot = \top, ~-\top = \bot$ (see more in the next section), we
obtain that $\R_{-} =  \{(f,f), (f,\bot), (\top, \bot), (f,\top),
(\bot,\top)\}$, while for the autoepistemic Moore's operator
\cite{Gins88}, $~o_i = \mu:\B \rightarrow \B$,  defined by\\ $\mu(x)
= t$ if $x \in \{\top, t\}$; $~f$ otherwise, we have that\\
 $\R_{\mu}= \{ (f, f), (f,\bot), (f,\top) (\bot,\top), (\top,\top)\}$.\\
Both of these modal operators are additive and normal. For the modal
negation additive operator $\widetilde{o_i} = \neg$, we have that\\
$\widetilde{\R_{\neg}} =  \{(f,f), (f,\bot), (f,\top),(\bot, \bot),
(\bot,f), (\top,\top), (\top,f) \}$.
\\$\square$\\
Now we are able to define the relational Kripke-style semantics for
a propositional modal logic $\L$, based on the modal Heyting
algebras in Proposition \ref{prop:canonic}:
\begin{definition} \label{def:kmodel} For a complete distributive lattice-based logics, the mapping $\mathbb{D}: \M \rightarrow \K_{\M}$ is defined as follows:
 Let  $(\textbf{A},I) \in \M $ be an algebraic model of $\L$, then $M_K = (K,I_K) = \mathbb{D}(\textbf{A},I)$ is the correspondent
  Kripke model, such that $K = \langle (1_K, \sqsubseteq),
  \{\R_j\}_{j \leq n}
\rangle$ is a frame, where $1_K = \widehat{A}$, $\R_j$   is an
accessibility relation (given by Definition \ref{def:frame}) for a
modal operator $o_j$, and $I_K:Var \rightarrow \P(A)$ is
 a canonical valuation, such that for any atomic formula (propositional
variable) $p \in Var$, $I_K(p) = \downarrow^+(I(p))~ \in D_K = A^+$.
Then, for any
world $x \in 1_K$, and formulae $\psi, \phi \in F(\L)$, \\
$M_K \models_x ~p ~~$ iff $~~x \in
I_K(p)$,\\
$M_K \models_x ~\phi \wedge \psi ~~$ iff $~~M_K \models_x ~\phi$
and $~~M_K \models_x ~\psi$,\\
$M_K \models_x ~\phi \vee \psi ~~$ iff $~~M_K \models_x ~\phi$
or $~~M_K \models_x ~\psi$,\\
$M_K \models_x ~\phi \Rightarrow \psi~~$ iff $~~ \forall y\in 1_K(
(y \sqsubseteq x ~$ and $ ~M_K \models_x
~\phi)$ implies  $~~M_K \models_y ~\psi)$,\\
$M_K \models_x ~\sim \phi~~$ iff $~~M_K \models_x ~\phi
\Rightarrow 0$,\\
 $M_K \models_x ~\diamondsuit_j
\phi~~$ iff $~~\exists y\in 1_K( (x,y) \in \R_j~$ and $ ~M_K
\models_y ~\phi )$, for each modal operator $o_j$,\\
$M_K \models_x ~\neg_j \phi~~$ iff $~~\forall y\in 1_K( ~M_K
\models_y ~\phi ~$ implies $~(x,y) \in \widetilde{\R_j}~)$,
for each negation modal operator $\widetilde{o_j}$.\\
 The mapping $\mathbb{E}:\K_{\M}
\rightarrow \M$ is defined as follows: for any $(K,I_K) \in
\K_{\M}$, $~~\mathbb{E}(K,I_K) = (\P(1_K),I_K) \in \M$.
\end{definition}
Notice that in the world $x = 0$ (bottom element in $A$) each
formula $\phi \in F(\L)$ is satisfied: because of that we will
denominate this world by inconsistent or \emph{trivial} world. The
semantics for the implication is the Kripke modal semantics for the
implication of the intuitionistic logic (only with inverted ordering
for the accessibility relation $\sqsubseteq$).\\ In any modal logic
the set of worlds where a formula $\phi$ is satisfied is denoted by
$\| \phi \| = \{ x~| ~M_K \models_x ~\phi \}$, so that we have $~M_K
\models_x ~\phi~$ iff $~ x \in \| \phi \|$.
 \begin{theo}\label{th:kmodel} \textsc{Soundness and Completeness:}
 Let  $(\textbf{A},I) \in \M $ be an algebraic model of $\L$ and
$(K,I_K) = \mathbb{D}(\textbf{A},I)$ be the correspondent
  Kripke model, with a frame
$K = \langle (1_K, \sqsubseteq), \{\R_j\}_{j \leq n} \rangle$, where
$1_K = \widehat{A}$, and the canonical valuation $I_K:Var\rightarrow
\P(A)$ given by Definition \ref{def:kmodel}. Than, for any
propositional
formula $\phi$, the set of worlds where $\phi$ holds is equal to \\
$~~~~~~\| \phi \| = \overline{I}_K(\phi) = i_n(\overline{I}(\phi))
~~\in D_K = A^+$,\\
where the monotone injection $i_n:A \hookrightarrow A_K$, $A_K =
\P(1_K)$, from Definition \ref{def:abs1}, satisfies $~~i_n =
\downarrow^+ $.
\end{theo}
\textbf{Proof:} By structural induction: \\
1. For any proposition variable $p \in Var$, $x \in \widehat{A}$,
$~M_K \models_x p~$ iff $~x \in I_K(p) = i_n \circ I(p) =
\downarrow^+
I(p)$, thus $\|p\| = \downarrow^+ I(p)$.\\
 2. From $M_K \models_x
\phi \wedge \psi~$ iff $~M_K \models_x \phi$ and $M_K \models_x
\psi$, holds that $\| \phi \wedge \psi\| = \| \phi\| \bigcap \| \psi
\|= \downarrow^+ \overline{I}(\phi) \bigcap \downarrow^+
\overline{I}( \psi)$, (by structural induction), $ = \downarrow^+
\overline{I}(\phi \wedge \psi)$(Prop.
\ref{def:canisomor}).\\
3.  Similarly, $\| \phi \vee \psi\| = \| \phi\| \bigcup \| \psi \| =
\downarrow^+ \overline{I}(\phi) \bigcup \downarrow^+ \overline{I}(
\psi) = \downarrow^+ \overline{I}(\phi
\vee \psi)$.\\
4. Suppose that $\| \phi \| = \downarrow^+ \overline{I}(\phi)$ and
$\|
\psi \| = \downarrow^+ \overline{I}(\psi)$. Then for any $x \in \widehat{A}$ we have that\\
$ x \in \| \phi \Rightarrow \psi \|$ iff $M_K \models_x ~\phi
\Rightarrow \psi~~$ iff $~~ \forall y\in \widehat{A}( (y \sqsubseteq
x~$ and $ ~M_K \models_y ~\phi)$ implies $~~M_K \models_y ~\psi)~~$
iff  $~~ \forall y\in \widehat{A}( (y \in \downarrow_A x~$ and $
~M_K \models_y ~\phi)$ implies $~~M_K \models_y ~\psi)~~$ iff (from
(*) holds $\downarrow^+x = \downarrow_A x$) $~~ \forall y( (y \in
\downarrow^+x \bigcap \|\phi\|$ implies $y \in \| \psi\|)~~$ iff
$~~\downarrow^+x \bigcap \|\phi\| \subseteq \| \psi\|$. So that $ S
=\| \phi \Rightarrow \psi \| =
\{x~|~\downarrow^+x \bigcap \|\phi\| \subseteq \| \psi\|\}$. \\
Then, $S = id_{A^+}(S) = \downarrow^+ \bigvee S = \bigcup
\downarrow^+S$ (from the homomorphism $\downarrow^+$) $ = \bigcup_{x
\in S} \downarrow^+ x = \bigcup \{\downarrow^+x~|~\downarrow^+x
\bigcap \|\phi\| \subseteq \| \psi\|\} = \bigcup \{S' \in A^+~|~S'
\bigcap \downarrow^+\overline{I}(\phi) \subseteq
\downarrow^+\overline{I}(\psi)\} = \downarrow^+(\overline{I}(\phi)
\rightharpoonup \overline{I}(\psi))$ (as shown in the example 5)
$=\downarrow^+\overline{I}(\phi \Rightarrow \psi)$ (from the
homomorphism of the valuation $\overline{I}$).\\ Consequently, $\|
\phi \Rightarrow \psi \|  = \downarrow^+\overline{I}(\phi
\Rightarrow \psi)$.\\
 5. For any additive algebraic modal operator
$o_i$ we obtain an existential logic modal operator
$\diamondsuit_i$, so that for any $x \in \widehat{A}$, $~M_K
\models_x \diamondsuit_i \phi/g~$ iff $~\exists y \in
\widehat{A}((x,y) \in \R_i$ and $M_K \models_y \phi/g)$,  iff
$~\exists y \in \widehat{A} ((x,y) \in \R_i$ and $y \in
\downarrow^+\alpha)~$,
 where $\alpha = \overline{I}(\phi/g)$.
Then, $~\|\diamondsuit_i \phi /g \| =\{x~|~\exists y ((x,y) \in
\R_i$ and $y \in \downarrow^+\alpha) \}~ = \\= \{x~|~\exists y  (y
\in \widehat{A} $ , $x \in \downarrow^+o_i(y)$ and $y \in
\downarrow^+\alpha) \} =\\ =~ \{x~|~y \in \downarrow^+\alpha$ and $x
\in \downarrow^+o_i(y) \}~ =~ \{x~|~y \in \downarrow^+\alpha$ and $x
\in \downarrow^+o_i \bigvee (\{y\}) \}~=~ \{x~|~y \in
\downarrow^+\alpha$ and $x \in o_i^+(\{y\}) \}~ = ~\bigcup_{y \in
\downarrow^+\alpha} o_i^+(\{y\})~ = ~$ (from the additivity of
$o_i$) $~ = ~o_i^+(\downarrow^+\alpha)~ =
~\downarrow^+o_i\bigvee(\downarrow^+\alpha)~ =
~~\downarrow^+o_i(\alpha)$.\\
Thus, we have that $~\|\diamondsuit_i \phi /g \| = \downarrow^+
o_i(\alpha) = \downarrow^+ o_i(\overline{I}(\phi /g )) =
\downarrow^+ \overline{I}(\diamondsuit_i \phi /g )$.\\
6. For any additive algebraic negation operator $\widetilde{o_i}$ we
obtain a  logic modal negation operator $\neg_i$, so that for any $x
\in \widehat{A}$, $~M_K \models_x ~\neg_i \phi~~$ iff $~~\forall y(
~M_K \models_y ~\phi $ implies $~(x,y) \in \widetilde{\R_i}~)$, iff
$~~\forall y( ~y \in \downarrow^+ \alpha~$ implies $~(x,y) \in
\widetilde{\R_i}~)$,
 where $\alpha = \overline{I}(\phi/g)$.
Then, $~\|\neg_i \phi /g \| =\{x~|~\forall y( ~y \in \downarrow^+
\alpha~$ implies $~(x,y) \in \widetilde{\R_i}~) \}~ \supseteq
$\\$\supseteq \downarrow^+ \widetilde{o_i} \alpha~~$
 (From definition of $\widetilde{\R_i}$ in Def.\ref{def:frame}).
 Let us show that also $~\|\neg_i \phi /g \| \subseteq \\ \downarrow^+ \widetilde{o_i}
 \alpha$. Suppose that there exists $x \in \widehat{A}$ (i.e., join-irreducible) such that $x
 \in \|\neg_i \phi /g \|$ but $ x \notin \downarrow^+ \widetilde{o_i}
 \alpha$. Then we define $\beta = \bigvee (\{x \} \bigcup \downarrow^+
 \widetilde{o_i}\alpha)$ (from Birkhoff Th. for distributive lattice each element is
  uniquely defined by the specific subset of join-irreducible
  elements).
 Thus (6.1) $\beta > \widetilde{o_i}
 \alpha$.\\
 Then for every $\gamma$ such that $\widetilde{o_i}\gamma \geq
 \beta$ (it always exists, at least for $\gamma = 0$, i.e., $\widetilde{o_i}\gamma = 1$) we have
 that $(\widetilde{o_i}\gamma, \gamma) \in \widetilde{\R_i}$, $(x,
 \gamma) \in \widetilde{\R_i}$. In order to have that $x
 \in \|\neg_i \phi /g \|$, i.e., $\forall y \in \downarrow^+ \alpha.
 (x,y) \in \widetilde{\R_i}$ it must hold that $\downarrow^+ \alpha
 \subseteq \downarrow^+ \gamma$. \\Then from $\widetilde{o_i}\alpha =
 (\bigvee \downarrow^+$ is an identity) $ =
 \widetilde{o_i}(\bigvee \downarrow^+ \alpha) = $ (from the additive
 property of the modal negation $\widetilde{o_i})  = \bigwedge
 \{\widetilde{o_i} y~|~y \in \downarrow^+ \alpha \} \geq \bigwedge
 \{\widetilde{o_i} y~|~y \in \downarrow^+ \gamma \} = \widetilde{o_i}(\bigvee \downarrow^+
 \alpha) = \widetilde{o_i}\gamma$, thus, $\widetilde{o_i}\alpha \geq
 \widetilde{o_i}\gamma \geq  \beta$ in contradiction with (6.1).
 Thus, we have that $~\|\neg_i \phi /g \| = \downarrow^+
\widetilde{o_i}(\alpha) = \downarrow^+
\widetilde{o_i}(\overline{I}(\phi /g )) = \downarrow^+
\overline{I}(\neg_i \phi /g )$.
\\ $\square$\\
 This theorem demonstrates that the satisfaction relation in
Definition \ref{def:kmodel} satisfies the general property for
relational semantics given by point 1 of Definition
\ref{def:newrelatsem}, that is, that holds $~~(K,m)
\models_{x}\phi~~$ iff  $~~x \in \overline{m}(\phi)$.\\ In fact, it
holds from the fact that for $m = I_K$, $~(K,m) \models_{x}\phi~$
iff $~x \in \| \phi \|$ and  from this theorem we have that $ \|
\phi \| =
\overline{m}(\phi)$.\\
 Notice
that in the case when a lattice $A$ is a complete ordering where for
any $x\in A$, $\downarrow^+x = \downarrow x $ (for example in the
fuzzy logic), then the minimum requirement for an unary modal
operators $o_i$ is to be monotonic.\\ We do not require it to be
surjective, by defining the accessibility relation as $\R_i
=\{(o_i(x),x)~|~x \in A\} \bigcup \{(x,0)~|~x \in A$ and $\nexists
y(x = o_i(y))\}$. In that case we have that $~M_K \models_x
\diamondsuit_i \phi~$ iff $~\exists y \in A.((x,y) \in \R_i$ and
$M_K \models_y  \phi)$, iff $~\exists y \in A.( x = o_i(y)$ and $M_K
\models_y  \phi)~$ iff (by inductive hypothesis $\|\phi\|
=\downarrow\overline{I}(\phi)$) $~\exists y \in A.(x = o_i(y)$ and
$y \sqsubseteq \downarrow \overline{I}(\phi))~$   iff (from the
monotonicity of $o_i$) $~\exists y \in A.(x = o_i(y) \sqsubseteq
o_i(\overline{I}(\phi))) ~$ iff $~\exists y \in A.(x = o_i(y) \in
\downarrow o_i(\overline{I}(\phi)))~$ iff (such $y$ exists, at least
as $0$) $~x \in \downarrow o_i(\overline{I}(\phi)) = \downarrow
(\overline{I}(\diamondsuit_i \phi))$.\\ Consequently,
$\|\diamondsuit_i
\phi \| = \downarrow (\overline{I}(\diamondsuit_i \phi)) = \downarrow^+ (\overline{I}(\diamondsuit_i \phi))$.\\
 Finally, from the canonical representation for distributive complete lattice based modal intuitionistic logics, we obtained that the
isomorphism, between the original  algebra $\textbf{A}$ with unary
modal operators and its canonical representation algebra
$\textbf{A}_K$, corresponds to the representation of  any
propositional formula $\phi$ by the set of worlds $\| \phi \|$ where
$\phi$ holds, in the canonical Kripke
model for the algebra $\textbf{A}$.\\
 So, for example, the term $\phi \wedge
\psi$ in the original algebra $(\textbf{A},I)$ corresponds to the
set $\| \phi \| \bigcap \| \psi \|$ in the canonical algebra
$(\textbf{A}_K, I_K)$, where $\| \phi \|$ is the set of worlds in
the canonical Kripke model  $( (\widehat{A}, \sqsubseteq),
\{\R_i\}, I_K )$  where $\phi$ holds.\\
As a consequence we obtained that this simple Kripke model is the
model of the normal modal logic with inference relation $~\psi
\vdash \phi~$ iff $~\|\psi\| \subseteq \|\phi\|$.\\ In fact, $~\psi
\vdash \phi~$ iff (based on the truth ordering) $~\overline{I}(\psi)
\sqsubseteq \overline{I}(\phi)~$ iff (based on the monotonicity of
$\downarrow^+$) $~\|\psi\| = \downarrow^+(\overline{I}(\psi))
\subseteq \downarrow^+ (\overline{I}(\phi)) = \|\phi\|~$.\\
 \section{Application to Belnap's bilattice}
 In this section we will apply the results obtained in the previous section to the 4-valued Belnap's bilattice
 based logic $\L$. Such a logic is a significant extension of normal strong Kleene's
 3-valued logic to the paraconsistent type of logics, where we are
 able to obtain a non-explosive inconsistency.\\ That is a very important
 class of logics which is able to deal also with
 mutually-inconsistent information, in typical Web data integration
 of different and independent source data with mutually inconsistent
 information \cite{MajkA04}. That is the main reason that we applied a new representation theorem to this case instead
 of more complex bilattices.\\
 Bilattice theory is a ramification of multi-valued logic by
considering both truth $\leq_t$ and knowledge $\leq_k$  partial
orderings. Given two truth values $x$ and $y$, if $x \leq_t y$ then
$y$ is at least as true as $x$, i.e., $x \leq_t y$ iff $x <_t y$ or
$x = y$. The negation operation for these two orderings, $\neg$ and
$-$ respectively, are defined as the involution operators which
satisfy De Morgan law between the join and meet operations.
\begin{definition} (Ginsberg ~\cite{Gins88})  \label{def:billat} A bilattice $\B$ is
defined as a sixtuple $(\B, \wedge, \vee,  \otimes, \oplus, \neg)$,
such that:  The t-lattice $(\B, \leq_t, \wedge, \vee)$ and the
k-lattice $(\B, \leq_k,\otimes, \oplus)$ are both complete lattices,
and  $\neg:\B \rightarrow \B$ is an involution ($\neg \neg$ is the
identity) mapping such that $\neg$ is a lattice homomorphism from
$(\B, \wedge, \vee)$ to $(\B, \vee, \wedge)$ and $(\B,\otimes,
\oplus)$ to itself.
\end{definition}
 The following definition introduces the subclass of D-bilattices \cite{Majk06Bi}
 (the Belnap's bilattice is the smallest non trivial D-bilattice). For
 more information and a more compact definition of D-bilattices and their properties,
 as well as a number of significant examples, the reader can use \cite{Majk07MV}.
\begin{definition}  \cite{Majk07MV} \label{def:d-billat}
A D-bilattice $\B$ is a  distributive bilattice $(\B, \wedge, \vee,
\otimes, \oplus,  \neg)$ with the isomorphism of truth-knowledge
lattices $~~\partial:(\B,\leq_t) \simeq (\B, \leq_k)$, which is an
involution. Let us define the unary operator $-~ =_{def}
\partial \neg \partial:\B \rightarrow \B$. Then we say
that a D-lattice  is \verb"perfect" if two truth negations, the
intuitionistic negation $\neg_t$ (pseudocomplement), such that
$\neg_t x = \bigvee \{z| z\wedge x = 0_t\}$, and the bilattice
negation $\neg$, are correlated by $~~\neg = \neg_t -$.
\end{definition}
In each  D-bilattice $(\B, \wedge, \vee, \otimes, \oplus, \neg)$,
the  operator $-$ is selfadjoint modal operator w.r.t. the $\leq_t$,
and   the bilattice negation operator for k-lattice satisfy $- 1_k
=0_k$, $- 0_k = 1_k$, while $- 1_t = 1_t$, $- 0_t = 0_t$.
\begin{coro} \cite{Majk07MV} \label{prop:d-billat}
For any D-bilattice $\B$ the duality operator $~\partial$ can be
extended to the following isomorphism of modal Heyting algebras \\
$~\partial:(\B,\leq_t, \alpha_t) \simeq (\B, \leq_k, \alpha_k)$,
with $\alpha_t = \{\wedge, \rightharpoonup, - \}$,  $\alpha_k  = \{
\otimes, \rightharpoondown, \neg \}$,\\ where $\rightharpoonup$ and
$\rightharpoondown$ are the intuitionistic implications (the
relative pseudocomplements) w.r.t. the $\leq_t$ and $\leq_k$
respectively.
\end{coro}
Informally, these dual lattices are  the modal extensions of Heyting
algebras. The conjugate modal operators are the \emph{belief}
operators.
As we will see, they correspond also to \emph{default} negations in
\emph{dual}
algebras. \\
The approach that we will use in order to find the representation
theorem for a Belnap's billatice (defined in Example 8),  based on
the fact that it is a D-bilattice, is different than the standard
one, based on the \emph{natural duality theorems} \cite{ClDa98}, (a
natural duality for a quasi-variety gives us a uniform method to
represent each algebra in the quasi-variety as the algebra of all
continuous homomorphisms over some structured Boolean space), but
close in spirit to the higher-order Herbrand model types
\cite{Majk06}.\\ A many-valued interpretation of a logic $\L$ in an
algebraic model $(\textbf{A},I ) = ((\B,\leq_k, \alpha_k), I)$ is of
the form $I:Var \rightarrow \B$, while for its extension
$(\textbf{A}_K, I_K)\\ = \mathbb{E}(\mathbb{D}((\textbf{A},I )))$
the interpretation is of the higher-order type $I_K:Var \rightarrow
A_K \subseteq \P(\B)\simeq \textbf{2}^{\B}$. That is, it maps each
propositional variable in $Var$ to a logical value which is a
\emph{function} $f$ in $\textbf{2}^{\B}$. Really, it maps to some
subset $S$ of $\B$, but such a set can be equivalently represented
by its characteristic function $f \in \textbf{2}^{\B}$, such that $S
= \{\alpha \in \B|f(\alpha) =1\}$. In what follows we will use both
of these
equivalent set-based and functional representations.\\
Both latices $(\B, \leq_t)$ and $(\B, \leq_k)$ are distributive
latices, thus, from the Proposition \ref{def:canisomor} we obtain
that
\begin{enumerate}
  \item For the truth-ordered lattice $(\B,\leq_t)$:
 $~~\B^+_t =
\{\downarrow^+ a ~| a \in \B~\} =\\
\{\{f\},\{f,\bot\},\{f,\top\},\{f,\bot,\top\}\} \subseteq
\P(\{f,\bot,\top\})$, with bottom $0_t = \downarrow^+ f = \downarrow
f = \{f\}$, and top element $1_t = \downarrow^+ t = \bigcup_{x\in
S_t = \{\bot, \top\}} \downarrow x =
\{f,\bot,\top\}$.\\
That is, we have the isomorphism $~i_t = \downarrow^+:(\B,\leq_t)
\simeq (\B^+_t, \subseteq)\subset (\P(1_t), \subseteq)$,
 such that  $i_t(f) = \{f\}, i_t(\bot) = \{f,\bot\}, i_t(\top) =
 \{f,\top\}$ and $i_t(t) = \{f,\bot,\top\}$, which satisfies the requirement (C)(ii) for inclusion  $i_n \equiv i_t$.
  \item For the knowledge-ordered lattice $(\B,\leq_k)$:
$~~\B^+_k = \{\downarrow^+ a ~| a \in \B~\} =\\
\{\{\bot\},\{\bot,f\},\{\bot,t\},\{\bot,f,t\}\} \subseteq
\P(\{\bot,f,t\})$, with bottom $0_k = \downarrow^+ \bot =\\
\downarrow \bot= \{\bot\}$, and top element $1_k = \downarrow^+ \top
= \bigcup_{x\in S_k = \{f, t \}} \downarrow x =
\{\bot,f,t\}$.\\
That is, we have the isomorphism $~i_k = \downarrow^+:(\B,\leq_k)
\simeq (\B^+_k, \subseteq)\subset (\P(1_k),\\ \subseteq)$,
 such that  $i_k(\bot) = \{\bot\}, i_k(f) = \{\bot,f\}, i_k(t) =
 \{\bot,t\}$ and $i_k(\top) = \{\bot,f,t\}$, which satisfies the requirement (C)(ii) for inclusion  $i_n \equiv i_k$.
 \end{enumerate}
 These  two lattices $(\B^+_{t},
 \subseteq)$ and $(\B^+_{k},
 \subseteq)$ satisfy the
 closure property \cite{McTa46} for elements of these lattices (from
 Proposition \ref{def:canisomor}), and we are able to define the
 relative-pseudocomplements for them (see the Example 5),
 $\rightharpoonup^+ = \downarrow^+ \rightharpoonup \bigvee$
 for $\B^+_t$ and $\rightharpoondown^+ = \downarrow^+ \rightharpoondown \bigvee$
 for $\B^+_k$. Thus, $(\B^+_t,
 \subseteq, \{\bigcap, \rightharpoonup^+\} )$ and $(\B^+_k,
 \subseteq, \{\bigcap,\rightharpoondown^+\} )$ are Heyting
 algebras. The negation is defined by $\neg_t X = X
 \rightharpoonup^+ 0_t$ for any $X \in \B^+_t$, and by $\neg_k X = X
 \rightharpoondown^+ 0_k$ for any $X \in \B^+_k$, respectively.\\
 But as Halmos has shown \cite{Halm62}, in the structures as $(\B^+_t, \subseteq)$
 (and also $(\B^+_k, \subseteq)$)  each closed element is
 also  open and can support also
 the  \emph{modal} operator $\diamond$ conjugate to itself.
 This is exactly our case.
 \begin{propo}  \label{prop:3} Let $\diamond_t$ and $\diamond_k$ be two operators on sets such that for a given
 set $X \in \P(1_t)$, $~\diamond_t X = \{-x~|~x \in X\}$, and  for $Y
 \in \P(1_k)$, $~\diamond_k Y =  \{\neg y~|~y \in Y\}$.
 Then $(\P(1_t), \subseteq, \{\bigcap, \rightharpoonup^+, \diamond_t\})$ and $(\P(1_k),
 \subseteq, \{\bigcap, \rightharpoondown^+, \diamond_k\} )$ are modal extensions of Heyting
 algebras.\\ Their restriction on $\B^+_t$ and $\B^+_k$ are $\diamond_t = \downarrow^+ - \bigvee$, $\diamond_k = \downarrow^+ \neg \bigvee$,
 and $(\B^+_t,
 \subseteq, \{\bigcap, \rightharpoonup^+, \diamond_t\})$ and $(\B^+_k,
 \subseteq, \{\bigcap, \rightharpoondown^+, \diamond_k\})$ are modal Heyting
 algebras.
 \end{propo}
 \textbf{Proof}: We have that $\diamond_t(\{f\}) = \{-f\} = \{f\}$,
 so $\diamond_t$ is normal modal operator, and, for any two sets $X,Y \in \P(1_t)$, $~\diamond_t(X
 \bigcup Y) = \{-x~|~x \in X
 \bigcup Y\} = \{-x~|~x \in X$ or $x \in Y\} = \{-x~|~x \in X\} \bigcup  \{-x~|~x \in Y\} = ~\diamond_t(X) \bigcup
 \diamond_t(Y)$, that is, $\diamond_t$ is additive.\\
 It is easy to show that for any $X \in \B^+_t$, $~\diamond_t X =
 \neg_t \diamond_t \neg_t X = \Box_t X \in \B^+_t$, thus $\diamond_t \equiv
 \Box_t$, that is, it is conjugate to yourself. The same holds for
 $\diamond_k$ w.r.t. $\B^+_k$, thus $(\B^+_t,
 \subseteq, \{\bigcap, \rightharpoonup^+, \diamond_t\})$ and $(\B^+_t,
 \subseteq, \{\bigcap, \rightharpoondown^+, \diamond_k\})$ are modal Heyting
 subalgebras of $(\P(1_t), \subseteq, \{\bigcap, \rightharpoonup^+, \diamond_t\})$ and $(\P(1_k),
 \subseteq, \{\bigcap, \rightharpoondown^+, \diamond_k\} )$ respectively.\\
 $\square$\\
 From Definition \ref{def:kmodel} and Theorem \ref{th:kmodel},
 for Kripke frames of these modal Heyting algebras we have that $K_t = (1_t,
\leq_t, R_-)$, where for the modal operator $\diamond_t$ the
accessibility relation is $\R_- = ~\{(x,y)~|~y  \in 1_t~$, and
$~x~\in \downarrow^+ -(y)\} = \{(f,f), (f,\bot), (\top, \bot),
(f,\top), (\bot, \top)\}$.\\
Dually, for knowledge ordering we obtain the Kripke frame $K_k =
(1_k, \leq_k, R_{\neg})$, where for a modal operator $\diamond_k$
the accessibility relation is $~~\R_{\neg} =\\~\{(x,y)~|~y  \in
1_k~$, and $~x~\in \downarrow^+ \neg(y)\}= \{(\bot, \bot), (\bot,f),
(t,
f), (\bot, t), (f,t)\}$.\\
It is easy to verify that these two Kripke frames are dual, i.e.,
$\partial_P:K_t \simeq K_k$.\\
Notice that we do not represent the bilattice negation $\neg$ as an
independent modal negation operator (in the truth-ordering lattice)
with an incompatibility relation (in Definition \ref{def:frame})
$\R_{\neg}$, because in Belnap's bilattice (see Example 7) it is
derived as the composition $\neg = - \neg_t = \neg_t -$ of the
selfadjoint (existential and universal) operator $-$ (conflation)
and pseudocomplement $\neg_t$. It is  represented as selfadjoint
modal operator in dual (knowledge ordering) lattice instead.
 Thus, for the propositional intuitionistic autoepistemic 4-valued logic $\L =
 (Var, \{\wedge, \Rightarrow, \flat\}, \Vvdash)$, where $\Rightarrow$
 is the intuitionistic implication and $\flat$ the  belief
 modal operator, we have:
\begin{theo} (Representation Theorem for Belnap's D-bilattice)\\
Let $~\partial:(\B,\leq_t, \alpha_t ) \cong (\B,\leq_k, \alpha_k)$
be a D-bilattice isomorphism for Belnap's bilattice $\B$, with $
\alpha_t = \{\wedge, \rightharpoonup, -\}$ and $ \alpha_k =
\{\otimes, \rightharpoondown, \neg\}$, and $I:Var \rightarrow \B$ be
a many-valued interpretation of intuitionistic autoepistemic logic
$\L = (Var, \{\wedge, \Rightarrow, \flat\}, \Vvdash)$. Let the
isomorphism $~\partial_{\P}~$ be the extension of the isomorphism
$\partial$ to sets, that is, for any set $X \in \P(1_t)$,
$~\partial_{\P}X = \{\partial x~|~x \in X\} \in \P(1_k)$, while
$~\partial^*_{\P}~$ be its reduction to $\B^+_t$ and $\B^+_k$
respectively.
 Then the following commutative diagram, where $I' =
\partial I$, $~I_t =
\downarrow^+_t I$, $~I_k = \downarrow^+_k \partial I$, $~\B^+_k =
\partial^*_{\P} (\B^+_t)$, $~1_k = \partial_{\P} (1_t)$, for algebraic
models of $\L$ holds $\vspace*{-3mm}$
\begin{diagram}
((\B,\leq_t, \alpha_t),I) & \rTo^{\partial} &  ((\B,\leq_k,\alpha_k),I')\\
\dTo^{\downarrow^+_t} & &\dTo_{\downarrow^+_k}\\
((\B^+_t,  \{\bigcap, \rightharpoonup^+, \diamond_t\}),I_t) &
\rTo^{\partial^*_{\P}} & ((\B^+_k,  \{\bigcap, \rightharpoondown^+, \diamond_k\}),I_k)\\
 \dTo^{i_t} & &\dTo_{i_k}\\
 ((\P(1_t),  \{\bigcap, \rightharpoonup^+, \diamond_t\}),I_t) &  \rTo^{\partial_{\P}}
      &((\P(1_k),   \{\bigcap, \rightharpoondown^+, \diamond_k\} ),I_k) \\
\dTo^{=} & &\dTo_{=}\\
(\mathbb{E}\circ \mathbb{D})((\B,\leq_t, \alpha_t),I)
&\rTo^{\partial_{\P}}       &
(\mathbb{E}\circ\mathbb{D})((\B,\leq_k,  \alpha_k),I')
\end{diagram}
where $in_t$, $in_k$ are injective homomorphisms, and
$\downarrow^+_t, \downarrow^+_k$ are the isomorphisms of
$\downarrow^+$ w.r.t the truth and knowledge ordering respectively.
\end{theo}
\textbf{Proof}: it is easy to verify, based on the precedent
propositions  \ref{def:canisomor}, \ref{prop:canonic}, \ref{prop:3},
and definition \ref{def:d-billat}. Let us consider a simple case,
for the term $\bot \wedge \top \in (\B,\leq_t, \{\wedge,
\rightharpoonup, -\})$. Then, $(\partial^*_{\P} \downarrow^+_t)(\bot
\wedge \top) =
\partial^*_{\P}(\{f,\bot\} \bigcap \{f,\top\}) =  \{\bot,f\}
\bigcap \{\bot,t\} = \{\bot\} = \downarrow^+_k( f \otimes t) =
(\downarrow^+_k
\partial)(\bot \wedge \top)$.\\ $\square$\\
 In this diagram we have to consider the
\emph{horizontal arrows} as a D-bilattice, from up to down: Belnap's
original D-bilattice, its set-based \emph{isomorphic
Representation}, and its powerset \emph{extension}. Notice that all
arrows (homomorphism between modal Heyting algebras) of the
commutative diagram on the top are \emph{isomorphisms}. The lower
part  of the commutative diagram represents the fact that the modal
Heyting algebras of isomorphic representations are the
\emph{subalgebras} of the powerset  extensions.
\section{Conclusion}

In this paper we defined a new framework for representation theorem,
based on models of a given many-valued modal logic $\L$ with
truth-invariance entailment, which is able to establish more close
link between algebraic and Kripke-style
  models for such non-classical logics.\\
The truth-invariance semantics of the entailment is different from
the matrix-based entailment, and, consequently,  this representation
theorem is substantially different from all previous representation
theorems with matrix-based
semantics.\\
 For the particular subclass of \emph{distributive} complete lattices we obtain  the possibility to define the canonical powerset extension algebra,
 based on the subsets of its carrier set of logic values, and its unique subalgebra
 isomorphic to the original many-valued algebra with modal
 operators.\\
 The resulting Kripke frame of the correspondent Kripke-style
 models of $\L$ has as the set of possible values the join-irreducible subset (with 0 element also)  of the carrier set of logic values
  of the many-valued algebra, in the way that we are able to represent the concrete Kripke models for a logic
 $\L$. Unlike the standard method based on the \emph{natural
 duality theorem} \cite{ClDa98}, where a class $\R$ of relational structures would
 be the family of duals of algebras,  difficult to describe in a
 simple logic language, our approach offers a very {simple and compact}
 autoreferential description. I believe that main results (representation theorem) can also be
obtained by Priestley duality in a different manner.
 The second contribution of this paper is
 dedicated to the representation theorem for Belnap's bilattice,
 which has recently been used for logic programs in Semantic Web applications \cite{MajkA04} in order to deal with incomplete
 and partially inconsistent information.

\bibliographystyle{IEEEbib}
\bibliography{mydb}

\end{document}